\documentclass[aps,twocolumn,amsfonts,showpacs]{revtex4-1}

\pdfoutput=1
\usepackage{epsfig}
\usepackage{psfrag}
\usepackage{amsmath}
\usepackage{amssymb}
\usepackage{color}
\usepackage{appendix}
\usepackage{hyperref}
\begin{document}
\author{Jia-Hao Su$^1$}
\author{Chuan-Yin Xia$^1$}
\author{Wei-Can Yang$^{1,2}$}
\author{Hua-Bi Zeng$^1$}\email{ hbzeng@yzu.edu.cn}

\affiliation{$^1$ Center for Gravitation and Cosmology, College of Physical Science
and Technology, Yangzhou University, Yangzhou 225009, China}
\affiliation{$^2$ Department of Physics, Osaka Metropolitan University, 3-3-138 Sugimoto, 558-8585 Osaka, Japan}

\title{Giant vortex in a fast rotating holographic superfluid}
\begin{abstract}

In a holographic superfluid disk, when the rotational velocity is large enough, we find  a giant vortex will form in the center of the system by merging several single charge vortices at some specific rotational velocity, with a phase stratification phenomenon for the order parameter.
The formation of a giant vortex can be explained as there is
not enough space for a standard vortex lattice.
Keep increasing the rotational velocity the giant vortex will disappear and there will be an appearance of a superfluid ring. In the giant vortex region, the number of vortices measured from winding number and
rotational velocity always satisfies the linear Feynman relation. However, when the superfluid ring starts to appear, the number of vortices  in the disk will decrease though the rotational velocity is increasing, where most of the order parameter is suppressed.

\end{abstract}
\maketitle
\section{introduction}

In quantum fluids, such as superfluid, a lot of interesting and counterintuitive phenomena occur, for example, in a rotating vessel containing superfluid, when the rotation velocity is large enough, topological defects with integer winding number, also known as quantum vortices, will be generated from the non-rotating superfluid \cite{vg1,vg2,vg3,vg4,vg5}.
Based on the requirement of minimum free energy, when the number of quantum vortices is enough, they will be arranged in the form of lattice \cite{VL1,VL2,VL3,VL4}. The vortices are always with winding number $n=1$, since a giant vortex with $n>1$ has larger free energy than a $n$ single charged vortices system\cite{mutiplechargestability}.

The earliest superfluid vortex lattices were observed in Helium-3 experiments \cite{ex1,ex2,ex3,ex4,ex5,ex6,ex7}. More research on the properties of vortex lattice occurs in numerical simulation and theoretical research, and the common method is to use the Gross-Pitaevskii(G-P) equation of mean field theory \cite{gp1,gp2,gp3,gp4,gp5,gp6}.
The Gross-Pitaevskii equation describes the ground state of a quantum system of identical bosons by using the Hartree-Fock approximation and the pseudopotential interaction model, which can well simulate various behaviors of the cold atomic BEC at zero temperature, including vortex lattice formation in rotational systems.
For a single component superfluid, it is generously believed that a hexagonal vortex lattice will be formed, however, the situation seems to be changed by fast rotation, which has been studied for more than a decade \cite{L,fast1,fast2,fast3,fast4,fast5}.
In the simulation of the G-P equation with a quadratic-plus-quartic potential \cite{quadratic-plus-quartic}, it is found that when the rotation velocity is large enough, the phase singularities will gather in the center and generate a "giant vortex"\cite{mine}.

Although the G-P equation is a powerful tool to study weakly interacting superfluid or BECs, it should be interesting to extend the study to the  holographic superfluid  that can naturally describe strongly interacting systems \cite{str1,sc1,sc3}. In addition, the finite temperature description of the G-P equation comes from the artificially added dissipation coefficient, while the holographic model comes from the natural dual of the black hole temperature \cite{3holo1,3h,3holo2}.
Therefore, we adopted the new method in view of the holographic model that can take strong interaction and temperature
into account to study the quantum vortex in superfluid .

At the end of last century, Maldacena's great conjecture that the two perspectives in string theory are equivalent brought holographic duality theory into public view \cite{str1}. Based on this equivalence, a complete set of duality models was constructed, namely, the equivalence of a high-dimensional gravitational field and a low-dimensional gauge field living on its boundary \cite{str2,str3}. In particular, when the large-N limit is considered, the gravitational field degenerates to classical gravity, while the gauge field has the property of strong interaction.
The holographic model has yielded stunning success in the study of strongly coupled systems.
This model can also be used to study condensed matter physics\cite{zaanen_liu_sun_schalm_2015,ammon_erdmenger_2015,hartnoll2018holographic}, such as holographic superconductivity, holographic superconductivity\cite{3holo1,3h,3holo2}.
There has been some previous studies of holographic superconductor and superfluid vortex\cite{HH1,HH2,HH3,HH4,HH5,HH6,HH7,HH8}, and the rotating superfluid dynamics evolution of vortices and Feynman relationship were studied at low rotation velocity \cite{lholo,Z,Feynman}. However, a research on the fast rotation holographic superfluid is still lacking.

In this paper, we use the finite temperature holographic superfluid model to study the vortex lattice properties of a fast rotating strongly coupled superfluid. We find that at some special rotational velocity, the disk center will generate a stable giant vortex structure with a multiple winding number, which violates the general understanding that the energy of multiple winding number vortices is higher than the energy of separated vortices\cite{mutiplechargestability}.
As the rotation velocity continues to increase, we observe the phase of giant vortex stratification in the radial direction on the phase configuration chart.
In an ultra-fast rotating superfluid, the vortices are distributed in rings, and the rings are isolated from each other. The outside part of it is a superfluid ring, and the inside part of it is a vortex ring.

The paper is organized as follows. In Sec. II we derive the fully expanded equation of motion of superfluid in the holographic model. In Sec. III we show the process of the time-dependent evolution at a specific rotational velocity where a giant vortex forms by the numerical solution of dynamic equations. And we obtain various configurations of vortices, including giant vortex and superfluid ring. In Sec. IV we discuss the relationship between quantity of vortices N and rotating velocity $\Omega$. We summarize our results in Sec. V.

\section{Holographic model and Equations of Motion for Time Evolution}

We construct the holographic superfluid model with a gauge field and a complex scalar field in the background of a planar
Schwarzschild  black hole in $3+1$ dimensional Anti-de Sitter spacetime, which dual to a $2+1$ dimensional conformal field theory on the boundary  \cite{3holo1,3h,3holo2}.
The action can be written as
\begin{equation}
S=\frac{1}{16\pi G_N} \int  d^4x \sqrt{-g}\Big[\mathcal{R}+6/L^2+\frac{1}{q^2}\mathcal{L}_{matter}
\Big]
\label{action}
\end{equation}
where $G_N$ is the Newton's constant, $\mathcal{R}$ is the Ricci scalar, $L$ is the radius of curvature of AdS spacetime.
 We work in the probe limit  by taking the large $q$ limit, which  means that the matter fields decouple from gravity.

The matter field lagrangian is
\begin{equation}
\mathcal{L}_{matter}=-\frac{1}{4}F_{\mu\nu}F^{\mu\nu}-|D_\mu\Psi|^2-m^2|\Psi|^2
\end{equation}
\label{lagrangian}
where $F_{\mu\nu}=\partial_\mu A_{\nu}-\partial_\nu A_{\mu}$ is the component of the $U(1)$ gauge field and $\Psi$ is complex scalar field with mass $m$. $D_\mu$ is the covariant derivative written as
\begin{equation}
D_\mu \Psi=\partial_\mu \Psi-iq_s A_\mu \Psi
\end{equation}

We choose the Eddington-Finkelstein coordinate $\sqrt{-g}=L^4/z^4$, which has form
\begin{equation}
\mathrm{d}s^2=L^2/z^2(-f(z)\mathrm{d}t^2-2\mathrm{d}t\mathrm{d}z+\mathrm{d}r^2+r^2\mathrm{d}\theta^2)
\end{equation}
$z=0$ represent the Ads boundary while $z=z_{h}$ is the horizon of black hole. Without loss of generality, we can set $z_{h}=L=1$, then $f(z)=1-z^3$, and Hawking temperature can be written as $T=3/4\pi$. The only characteristic parameter of the holographic superfluid is the dimensionless ratio $\mu/T$, where $\mu=A_t(z=0)$ is the chemical potential on the boundary. Then, the  temperature  on  the  boundary  can  be  expressed  as $T=(\mu_c/\mu)T_c$.

Then we can build polar coordinates in a disk on the dual 2 + 1 dimensional boundary to study the rotating superfluid.
In this model, the black hole is non rotation, thus the superfluid is treated as static but the disk is rotating, which also has a relative rotation.
By solving the action, we obtain the equation of motion of the scalar field
\begin{equation}
D^\mu D_\mu \Psi-m_s^2 \Psi=0
\end{equation}
and the equation of the vector field
\begin{equation}
\partial^\nu F_{\nu\mu}-iq_s(\Psi^*D_\mu\Psi-\Psi D_\mu\Psi^*)=0
\end{equation}
here, according to the conservation principle,  $J^\mu=i(\Psi^*D_\mu\Psi-\Psi D_\mu\Psi^*)$ is the bulk current.

By taking $m_s^2=-2$, the conformal dimension of the scalar field are $\Delta=\frac{3}{2}\pm \sqrt{\frac{9}{4}-m^2L^2}=\frac{3}{2} \pm \frac{1}{2}$, which means the expansion of the solution of scalar field and gauge field near the AdS boundary has the form
\begin{eqnarray}
\Psi = \phi z + \psi z^2 + \mathcal{O}(z^3)\\
A_\nu =a_\nu +b_\nu z + \mathcal{O}(z^2)
\end{eqnarray}
At the Ads conformal boundary $z=0$, we close the source $\psi|_{z=0}=0$. $a_t=\mu$ is the chemical potential and $b_t=\rho$ the charge. As we has discussed above, $\mu$ is inversely proportional to temperature $T$, when $\mu$ is exceeds a critical value $\mu_c=4.07$, the system spontaneously breaks the U(1) gauge symmetry and the expectation value of scalar operator $\left \langle O \right \rangle =\phi |_{z=0}$  has a finite-valued solution. In this paper, we fixed $\mu=6$ and the disk size radius R=5.

$a_\theta$ and $a_r$ represent the superfluid velocity, then we can add a rotation in the system without radial flow $a_r=0$
\begin{equation}
A_\theta|_{z=0}=a_\theta=\Omega r^2
\end{equation}
$\Omega$ is the rigid rotation with angular velocity  of the disk \cite{HH6}.

The fully expanded equation of motion can be written as
\begin{equation}
\begin{split}
-\left[\partial_{z}\left(f\partial_{z}\Psi\right)+i\left(\partial_{z}A_{t}\right)\Psi+2iA_{t}\partial_{z}\Psi\right]\\
+\left[-\partial_{r}^{2}\Psi+i\left(\partial_{r}A_{r}\right)\Psi+2iA_{r}\partial_{r}\Psi-\frac{1}{r}\partial_{r}\Psi\right]\\
+\frac{1}{r^{2}}\left[-\partial_{\theta}^{2}\Psi+i\left(\partial_{\theta}A_{\theta}\right)\Psi+2iA_{\theta}\partial_{\theta}\Psi\right]\\
+\left[A_{r}^{2}+\frac{iA_{r}}{r}+\frac{A_{\theta}^{2}}{r^{2}}+z\right]\Psi+2\partial_{t}\partial_{z}\Psi =0
\end{split}
\end{equation}

\begin{equation}
\begin{split}
\partial_{z}^{2}A_{t}-\partial_{z}\partial_{r}A_{r}-\frac{1}{r}\partial_{z}A_{r}-\frac{1}{r^{2}}\partial_{z}\partial_{\theta}A_{\theta}\\
-i\left(\Psi^{*}\partial_{z}\Psi-\Psi\partial_{z}\Psi^{*}\right)=0
\end{split}
\end{equation}

\begin{equation}
\begin{split}
-\frac{1}{r}\partial_{z}A_{r}-\partial_{t}\partial_{z}A_{t}-\partial_{t}\partial_{r}A_{r}-\frac{1}{r^{2}}\partial_{t}\partial_{\theta}A_{\theta}+\frac{f}{r}\partial_{z}A_{r}\\
+f\partial_{z}\partial_{r}A_{r}+\frac{f}{r^{2}}\partial_{z}\partial_{\theta}A_{\theta}+\partial_{r}^{2}A_{t}+\frac{1}{r}\partial_{r}A_{t}+\frac{1}{r^{2}}\partial_{\theta}^{2}A_{t}\\
-i\left(\Psi^{*}\partial_{t}\Psi-\Psi\partial_{t}\Psi^{*}\right)-2A_{t}\Psi^{*}\Psi\\
+if\left(\Psi^{*}\partial_{z}\Psi-\Psi\partial_{z}\Psi^{*}\right)=0
\end{split}
\end{equation}

\begin{equation}
\begin{split}
2\partial_{t}\partial_{z}A_{r}-\partial_{z}\partial_{r}A_{t}-\partial_{z}\left(f\partial_{z}A_{r}\right)+\frac{1}{r^{2}}\left(\partial_{r}\partial_{\theta}A_{\theta}
-\partial_{\theta}^{2}A_{r}\right)\\
+i\left(\Psi^{*}\partial_{r}\Psi-\Psi\partial_{r}\Psi^{*}\right)+2A_{r}\Psi^{*}\Psi=0
\end{split}
\end{equation}

\begin{equation}
\begin{split}
2\partial_{t}\partial_{z}A_{\theta}-\partial_{z}\left(f\partial_{z}A_{\theta}\right)-\partial_{z}\partial_{\theta}A_{t}-\partial_{r}^{2}A_{\theta}+\frac{1}{r}\partial_{r}A_{\theta}\\
+\partial_{r}\partial_{\theta}A_{r}-\frac{1}{r}\partial_{\theta}A_{r}+i\left(\Psi^{*}\partial_{\theta}\Psi-\Psi\partial_{\theta}\Psi^{*}\right)+2A_{\theta}\Psi^{*}\Psi=0
\end{split}
\end{equation}

For the numerical simulation, Chebyshev spectral methods are used in the (z,r) direction. Fourier spectral methods are used in the $\theta$ direction. Time evolution is simulated by the fourth-order Runge-Kutta method. The initial configuration at t=0 is chosen to be a homogenous superfluid state without any rotation at a fixed temperature below $T_c$. The temperature is $T\approx\frac{2}{3}T_c$ .

\section{THE STRUCTURES OF GIANT VORTEX AND SUPERFLUID RING}

\begin{figure}[h]
\begin{minipage}{0.22\linewidth}
 	\vspace{0pt}
 	\centerline{\includegraphics[trim=2.3cm 9.0cm 3.8cm 9.7cm, clip=true, scale=0.16, angle=0]{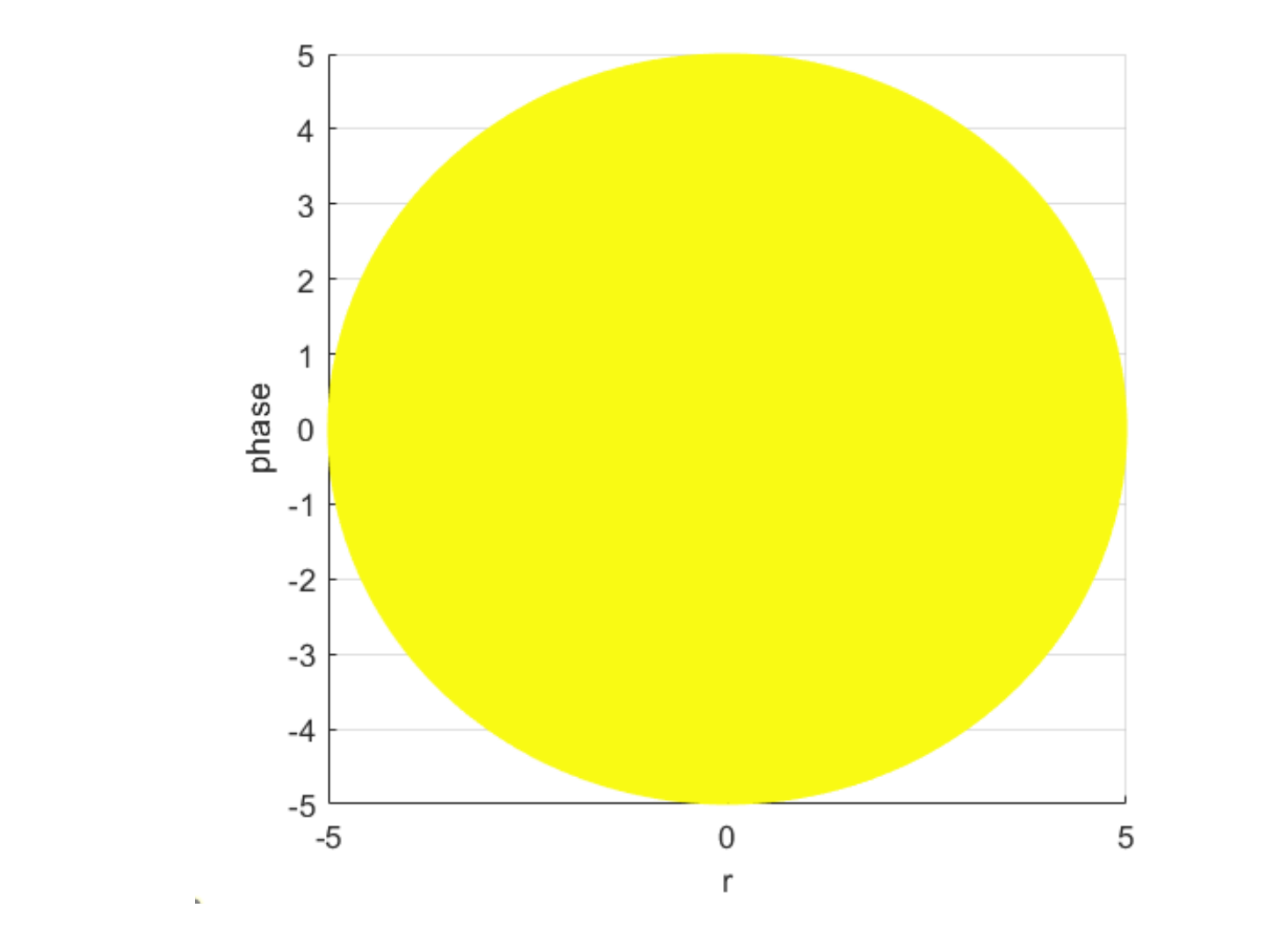}}
 	\vspace{0pt}
 	\centerline{\includegraphics[trim=2.3cm 9.0cm 3.8cm 9.7cm, clip=true, scale=0.16, angle=0]{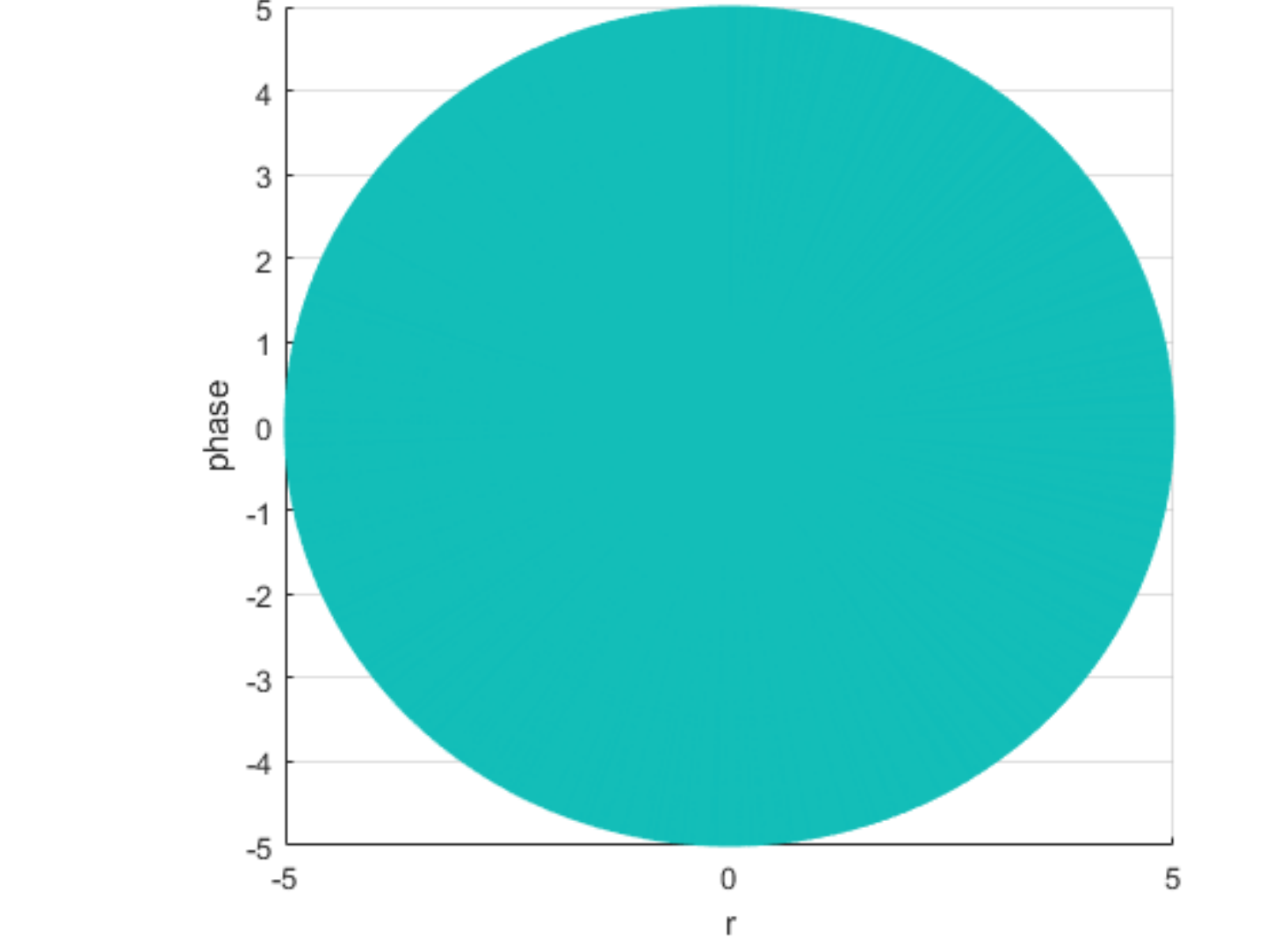}}
 	\vspace{0pt}
 \centerline{t=0}
 \end{minipage}
 \begin{minipage}{0.22\linewidth}
 	\centerline{\includegraphics[trim=2.3cm 9.0cm 3.8cm 9.7cm, clip=true, scale=0.16, angle=0]{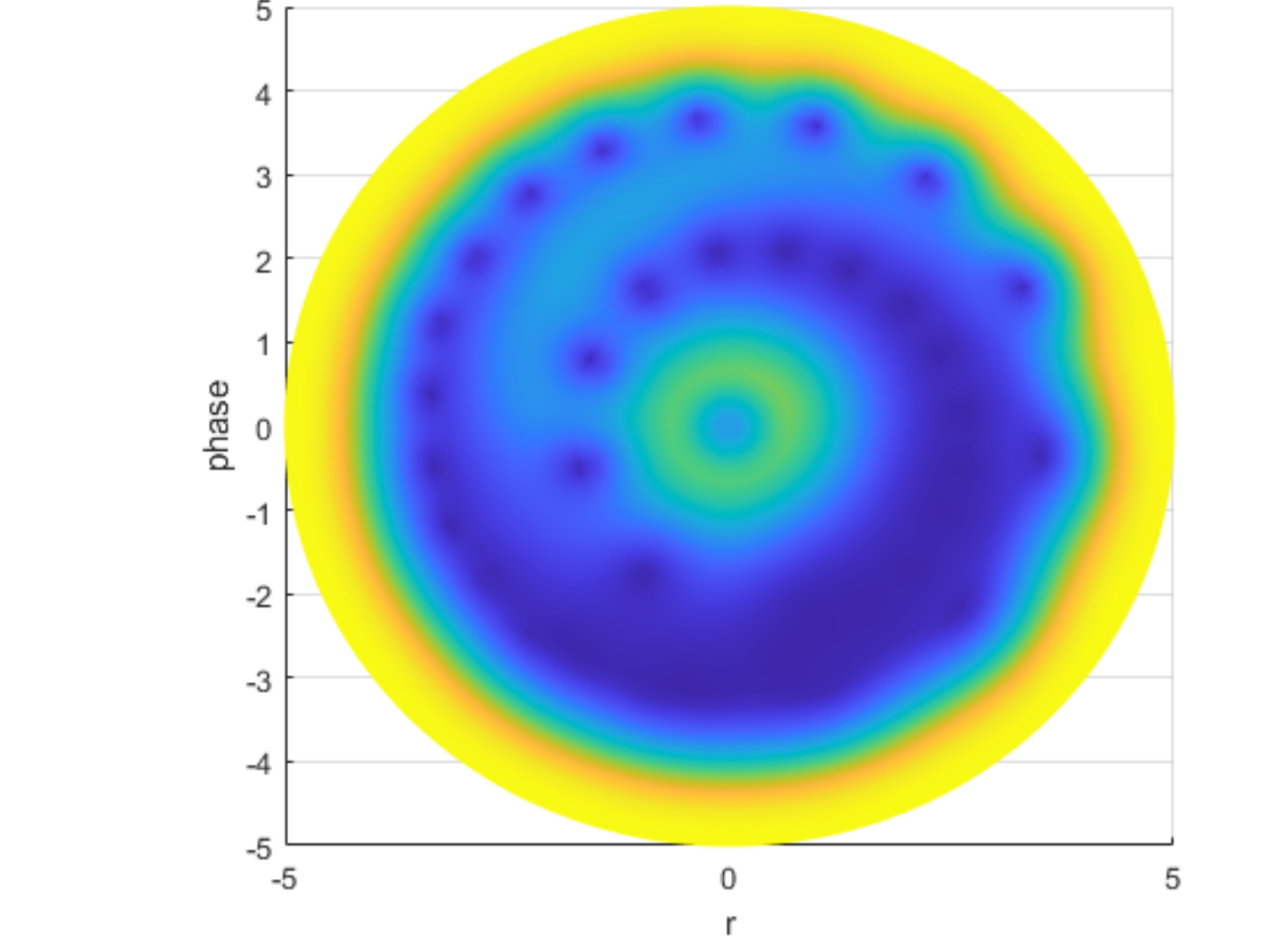}}
 	\vspace{0pt}
 	\centerline{\includegraphics[trim=2.3cm 9.0cm 3.8cm 9.7cm, clip=true, scale=0.16, angle=0]{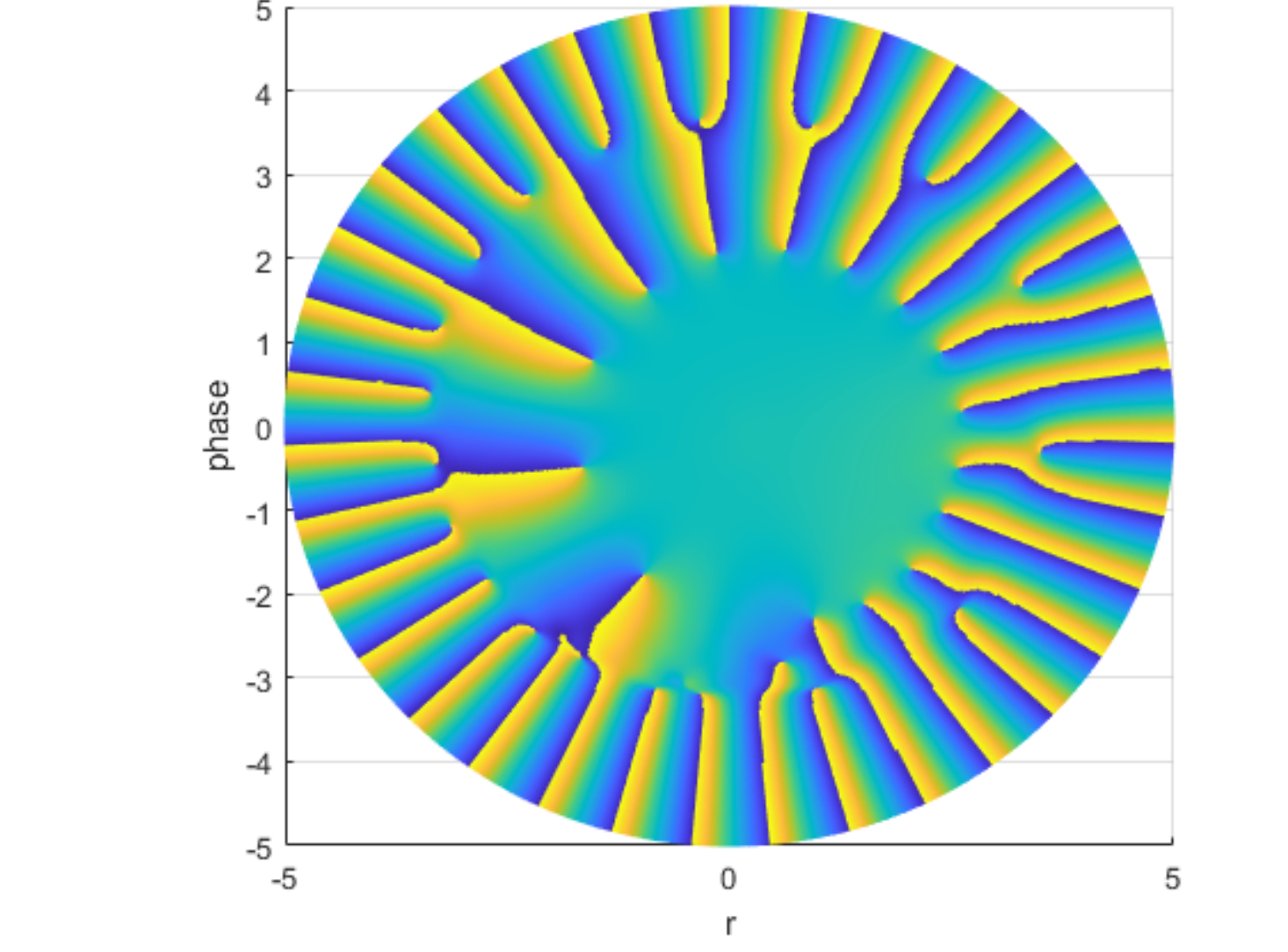}}
 	\vspace{0pt}
 	\centerline{t=500}
 \end{minipage}
 \begin{minipage}{0.22\linewidth}
	\vspace{0pt}
	\centerline{\includegraphics[trim=2.3cm 9.0cm 3.8cm 9.7cm, clip=true, scale=0.16, angle=0]{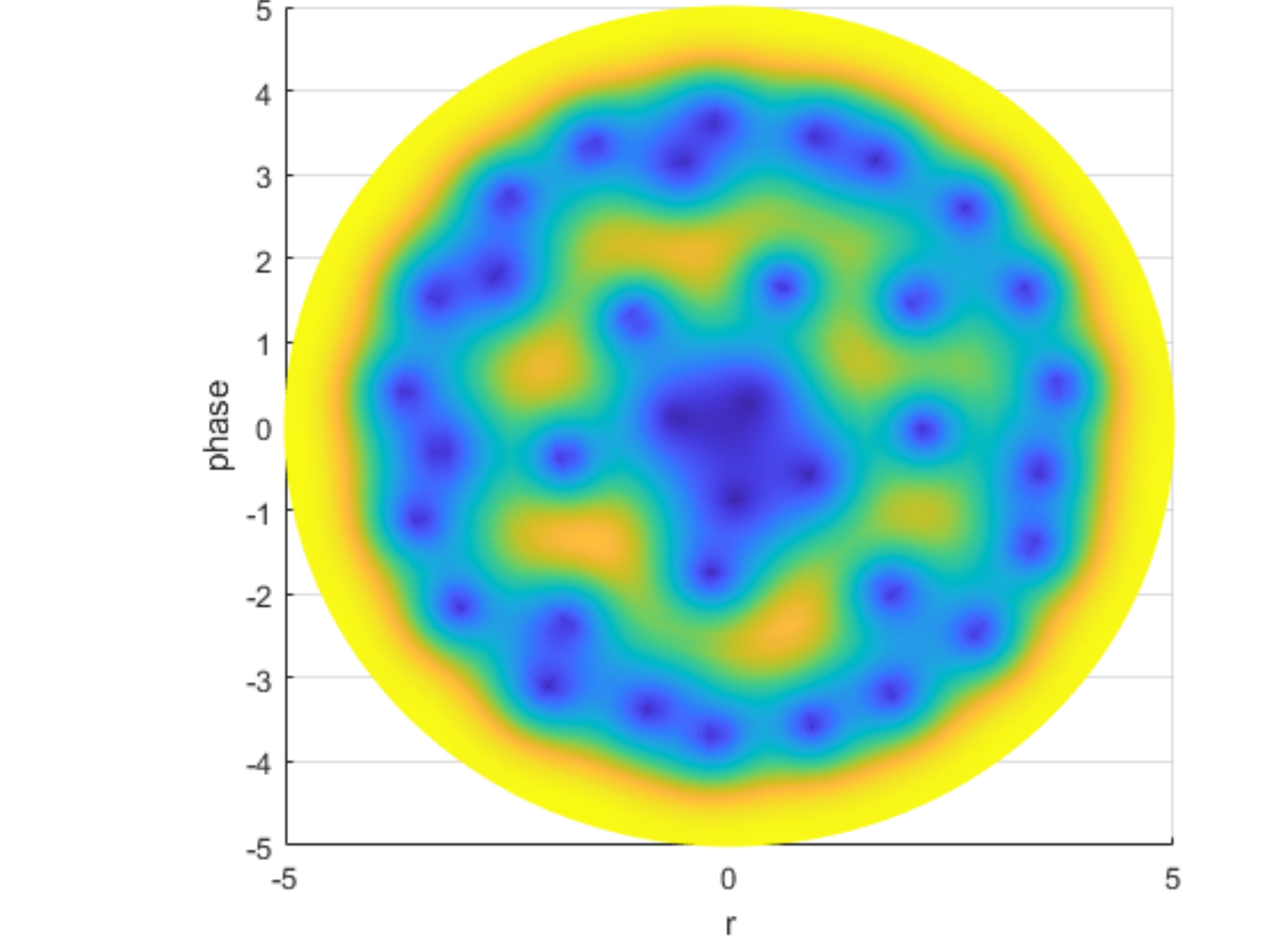}}
	\vspace{0pt}
	\centerline{\includegraphics[trim=2.3cm 9.0cm 3.8cm 9.7cm, clip=true, scale=0.16, angle=0]{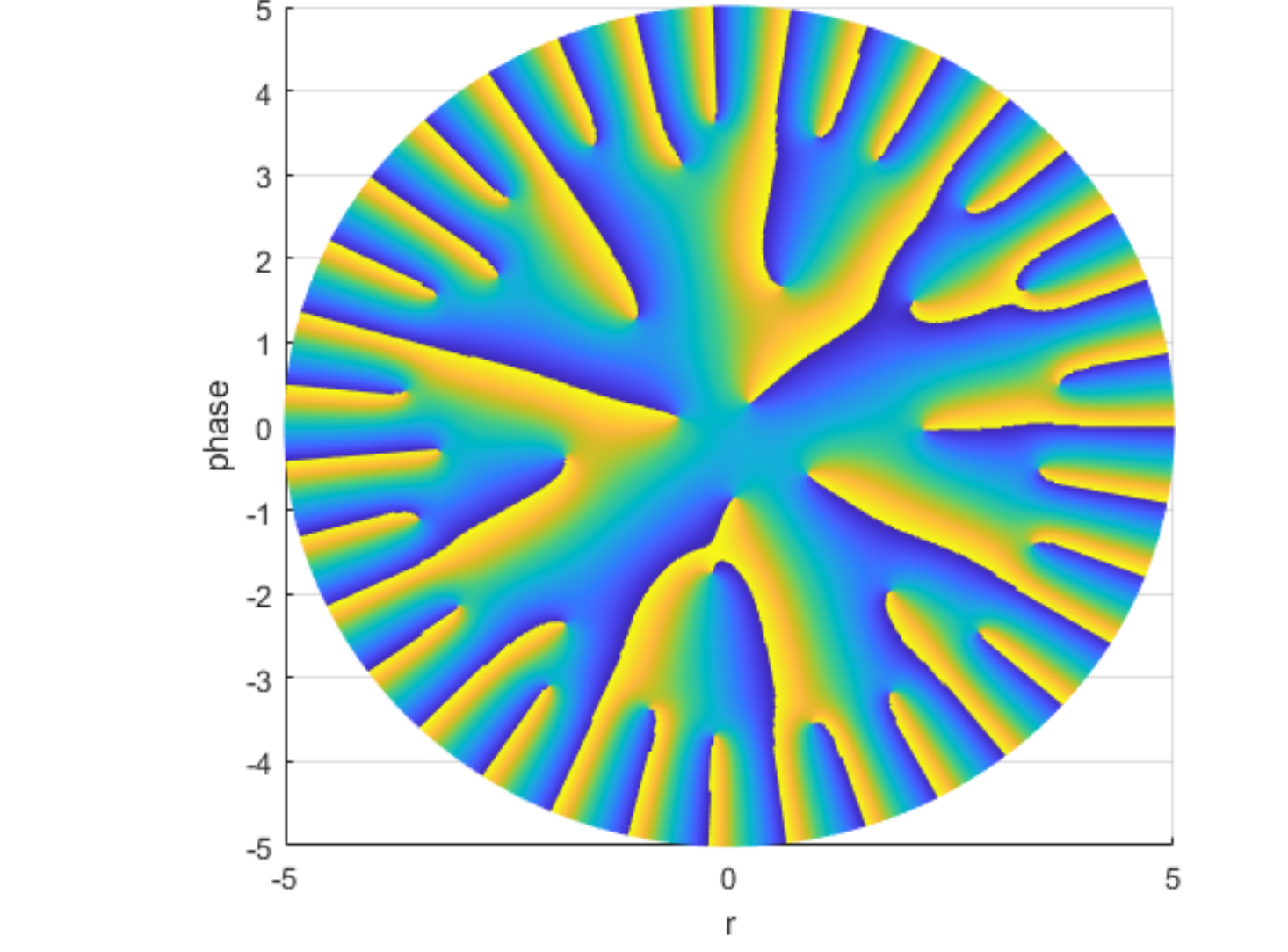}}
	\vspace{0pt}
\centerline{t=1000}
\end{minipage}
\begin{minipage}{0.22\linewidth}
	\centerline{\includegraphics[trim=2.3cm 9.0cm 3.8cm 9.7cm, clip=true, scale=0.16, angle=0]{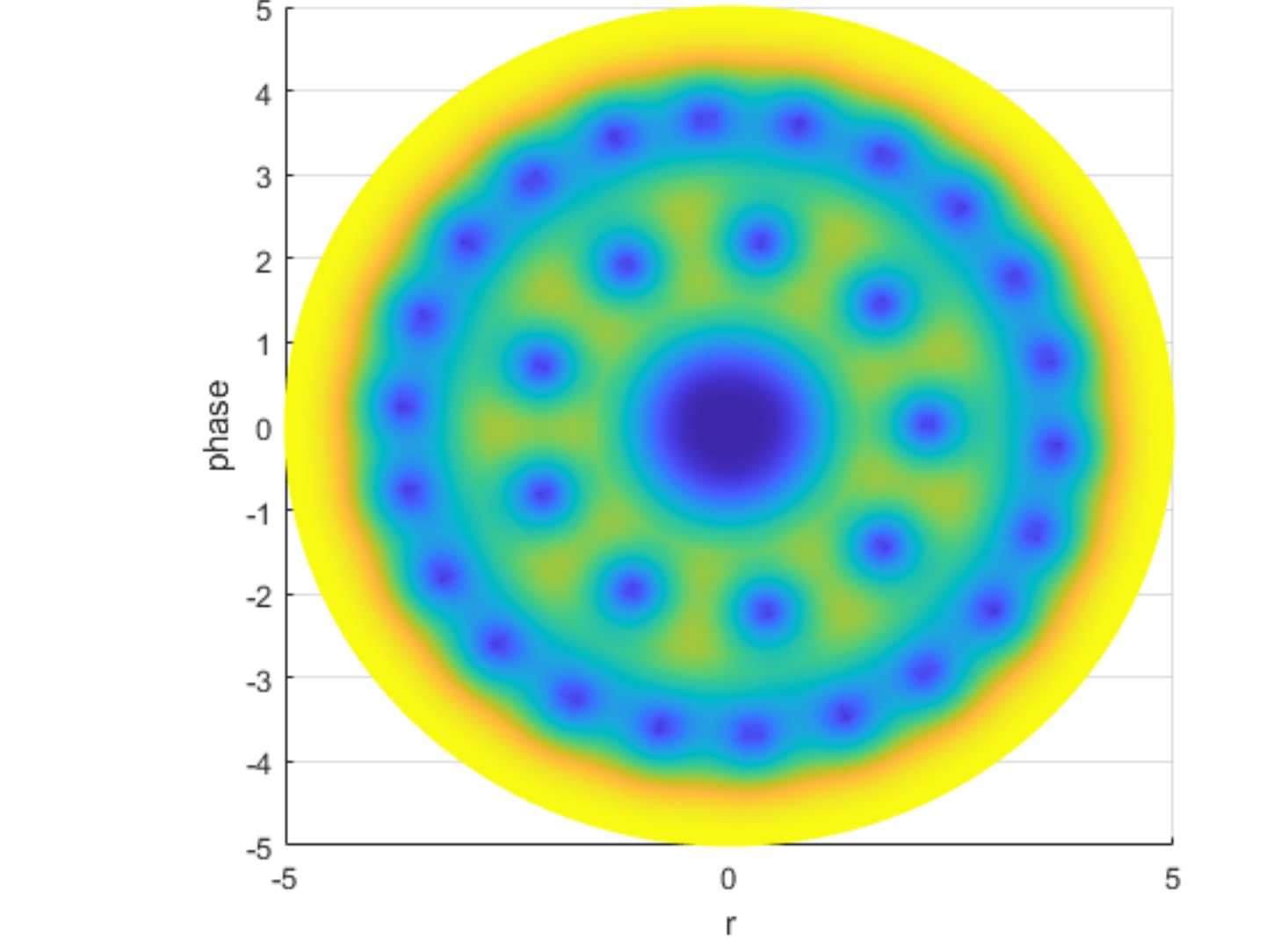}}
	\vspace{0pt}
	\centerline{\includegraphics[trim=2.3cm 9.0cm 3.8cm 9.7cm, clip=true, scale=0.16, angle=0]{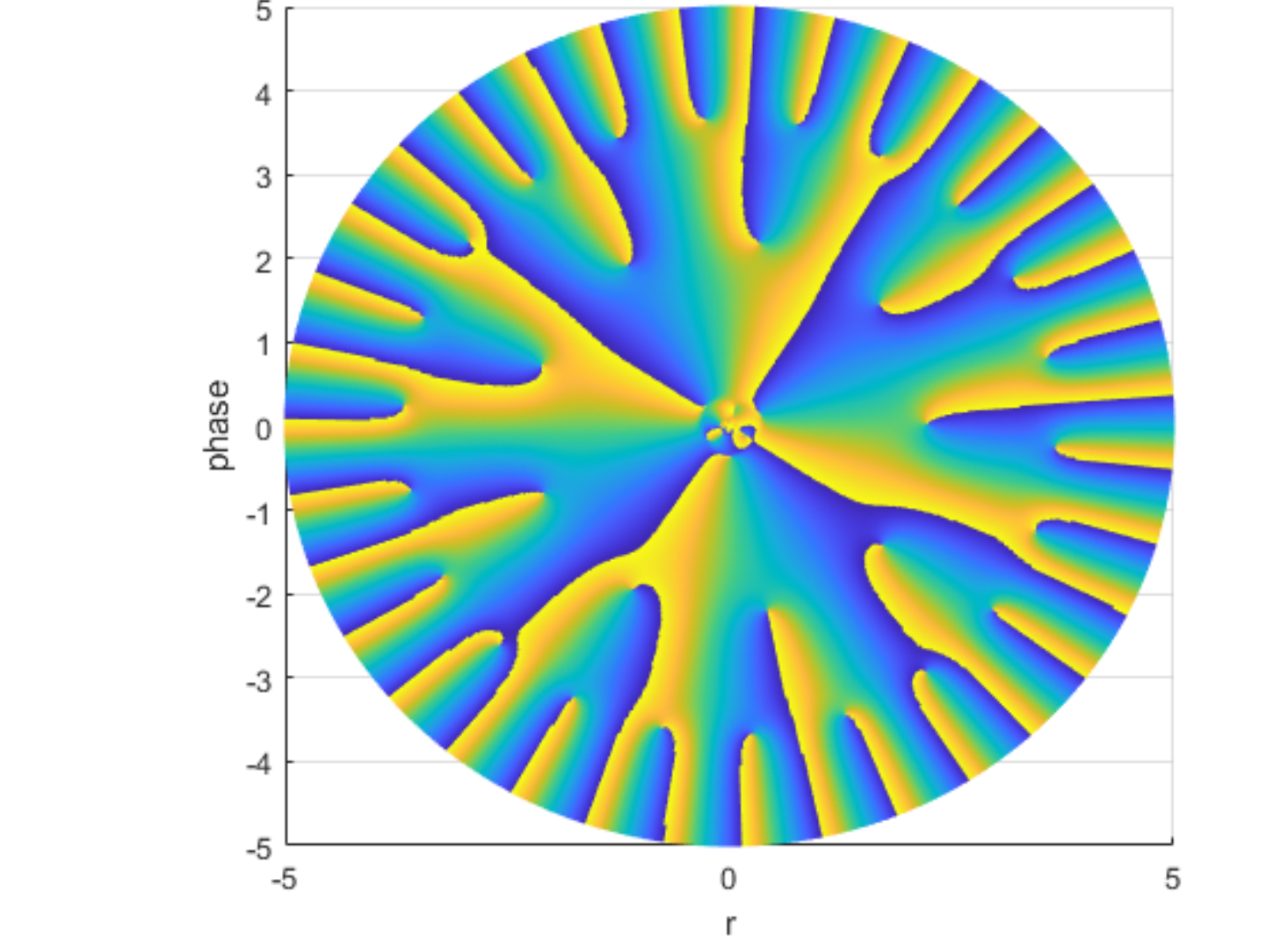}}
	\vspace{0pt}
	\centerline{t=2000}
\end{minipage}
\caption{This exhibits the formation of a four winding number giant vortex at $\Omega$=1.65, R=5. The uppers show the order parameter and the bottoms show the phase configuration in system at the corresponding time. }\label{fig1}
\end{figure}

After we rotate the disk superfluid, vortices will gradually enter from the boundary. An example of the dynamic formation process of a giant vortex in a rapidly rotating superflow  is shown in Fig.\ref{fig1}. Initially, the system was homogeneous. Then, when we give an angular velocity to the disk, the vortices emerge on the boundary, they move inside and form the formation of a giant vortex appearing in a two-layer structure. There are four vortices arriving at the center that produce a giant vortex which is not supposed to appear in a slow rotation case.

When the rotation velocity $\Omega$ ranges from $0.345$ to $0.45$, which is the slow rotation case, and the vortics lattice of polygonal grid\cite{gp1,gp2,gp3,gp4,gp6} can be obtained both in the G-P equation of mean field theory and holographic model.

\begin{figure}[ht]
\begin{minipage}{0.38\linewidth}
\includegraphics[trim=4.3cm 9.0cm 3.8cm 9.7cm, clip=true, scale=0.25, angle=0]{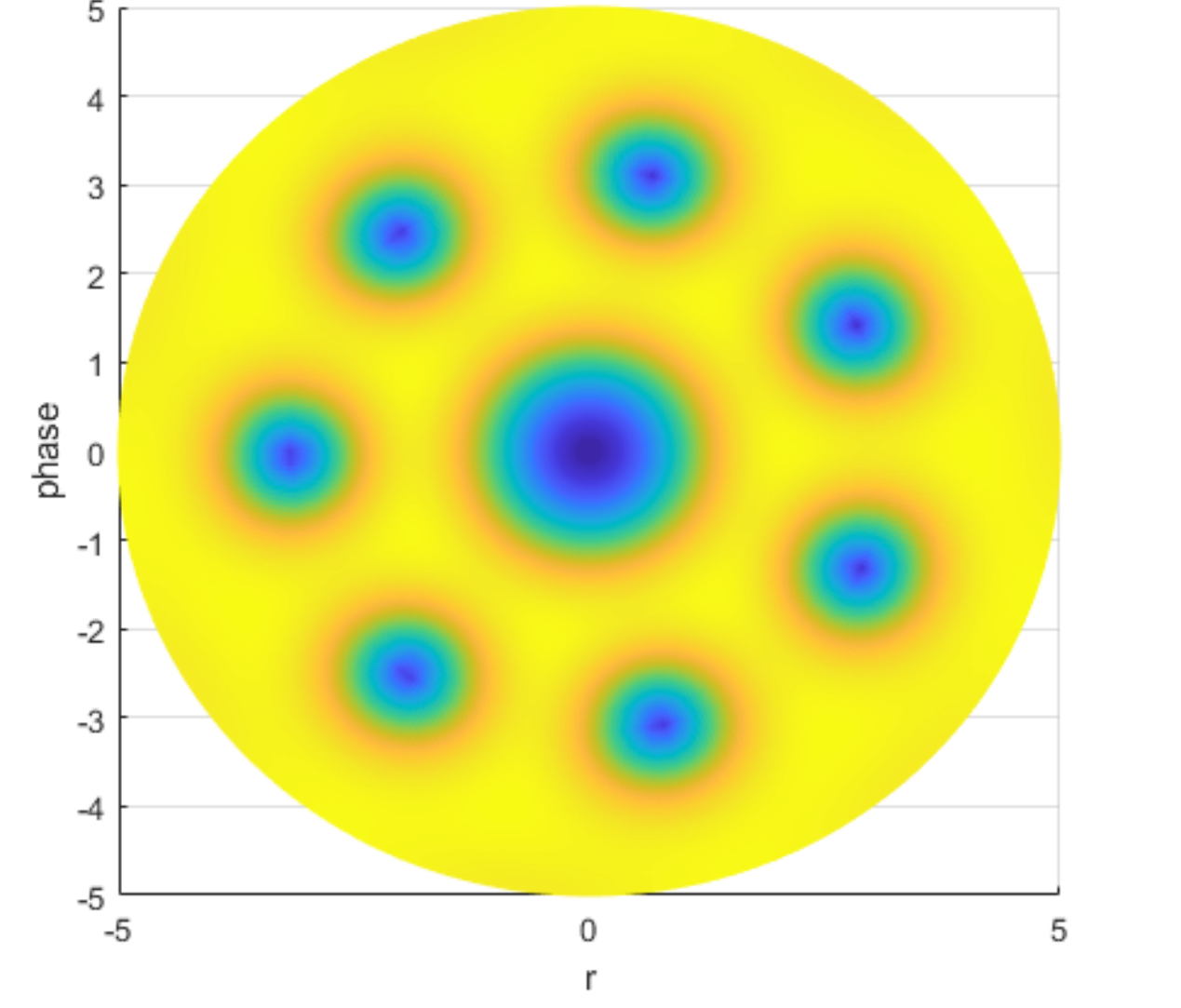}
\centerline{a}
\end{minipage}
\hspace{0mm}
\begin{minipage}{0.38\linewidth}
\includegraphics[trim=3.3cm 9.0cm 3.8cm 9.7cm, clip=true, scale=0.25, angle=0]{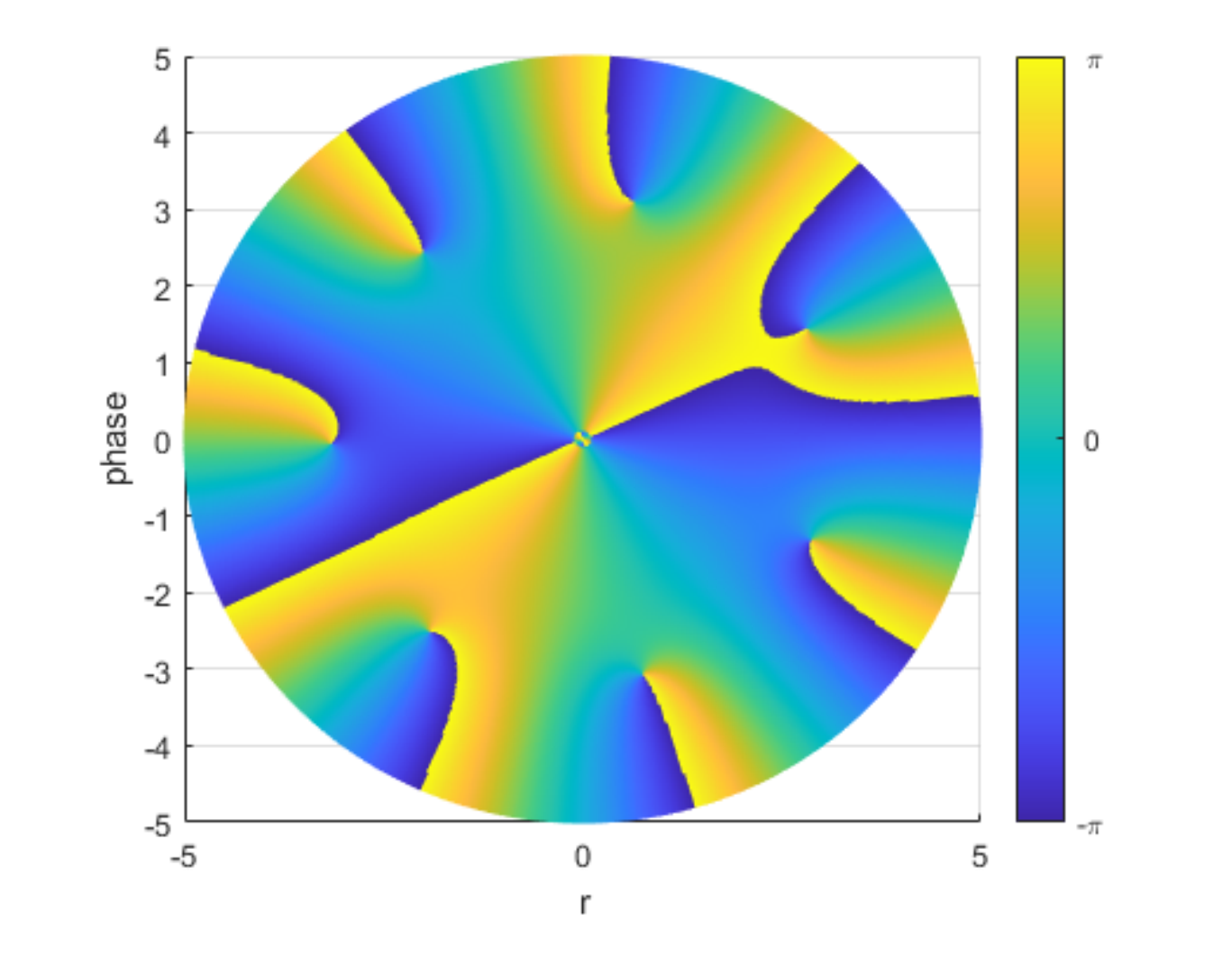}
\centerline{b}
\end{minipage}
\hspace{0mm}
\begin{minipage}{0.38\linewidth}
\includegraphics[trim=4.3cm 9.0cm 3.8cm 9.7cm, clip=true, scale=0.25, angle=0]{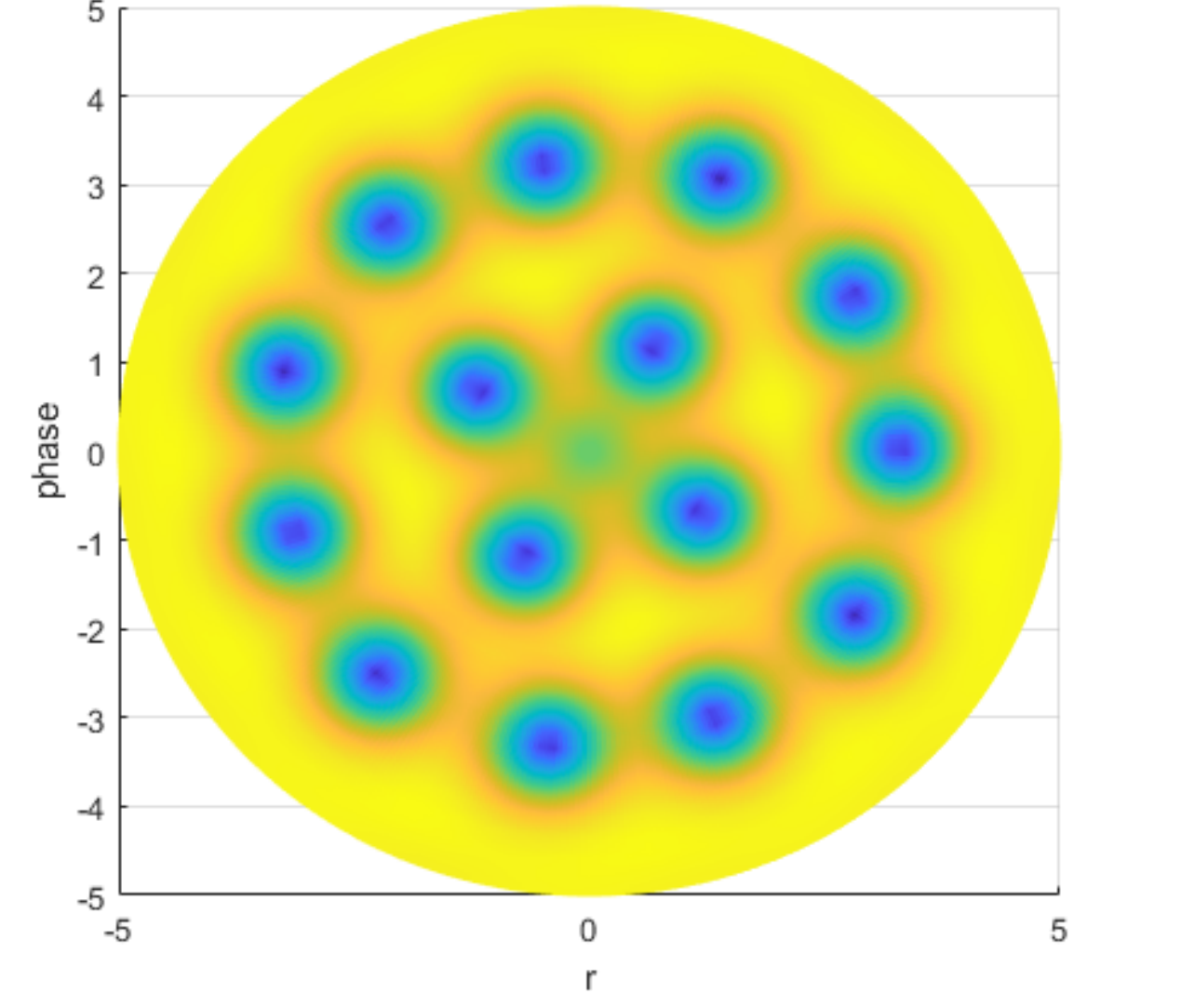}
\centerline{c}
\end{minipage}
\hspace{0mm}
\begin{minipage}{0.38\linewidth}
\includegraphics[trim=3.3cm 9.0cm 3.8cm 9.7cm, clip=true, scale=0.25, angle=0]{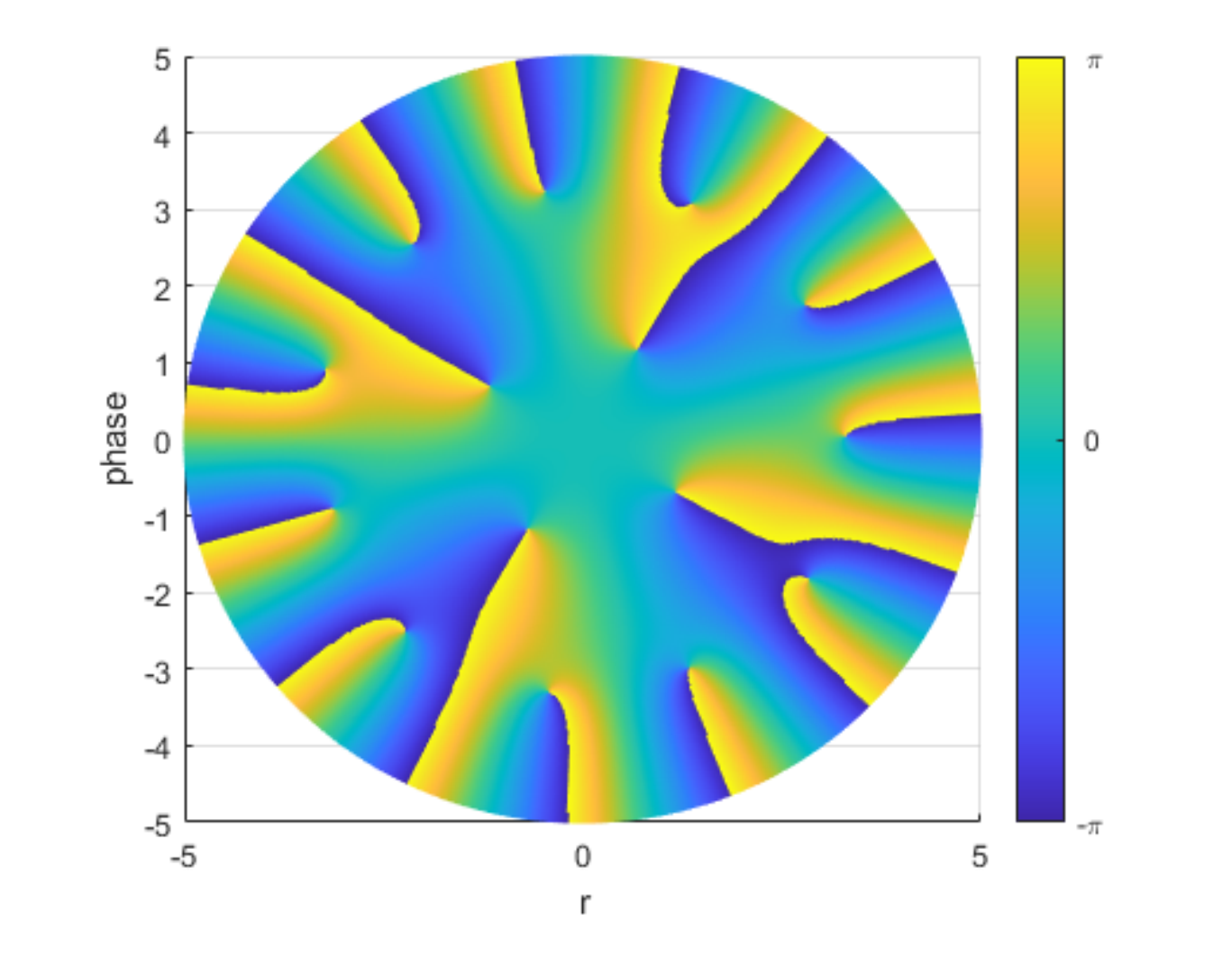}
\centerline{d}
\end{minipage}
\hspace{0mm}
\begin{minipage}{0.38\linewidth}
\includegraphics[trim=4.3cm 9.0cm 3.8cm 9.7cm, clip=true, scale=0.25, angle=0]{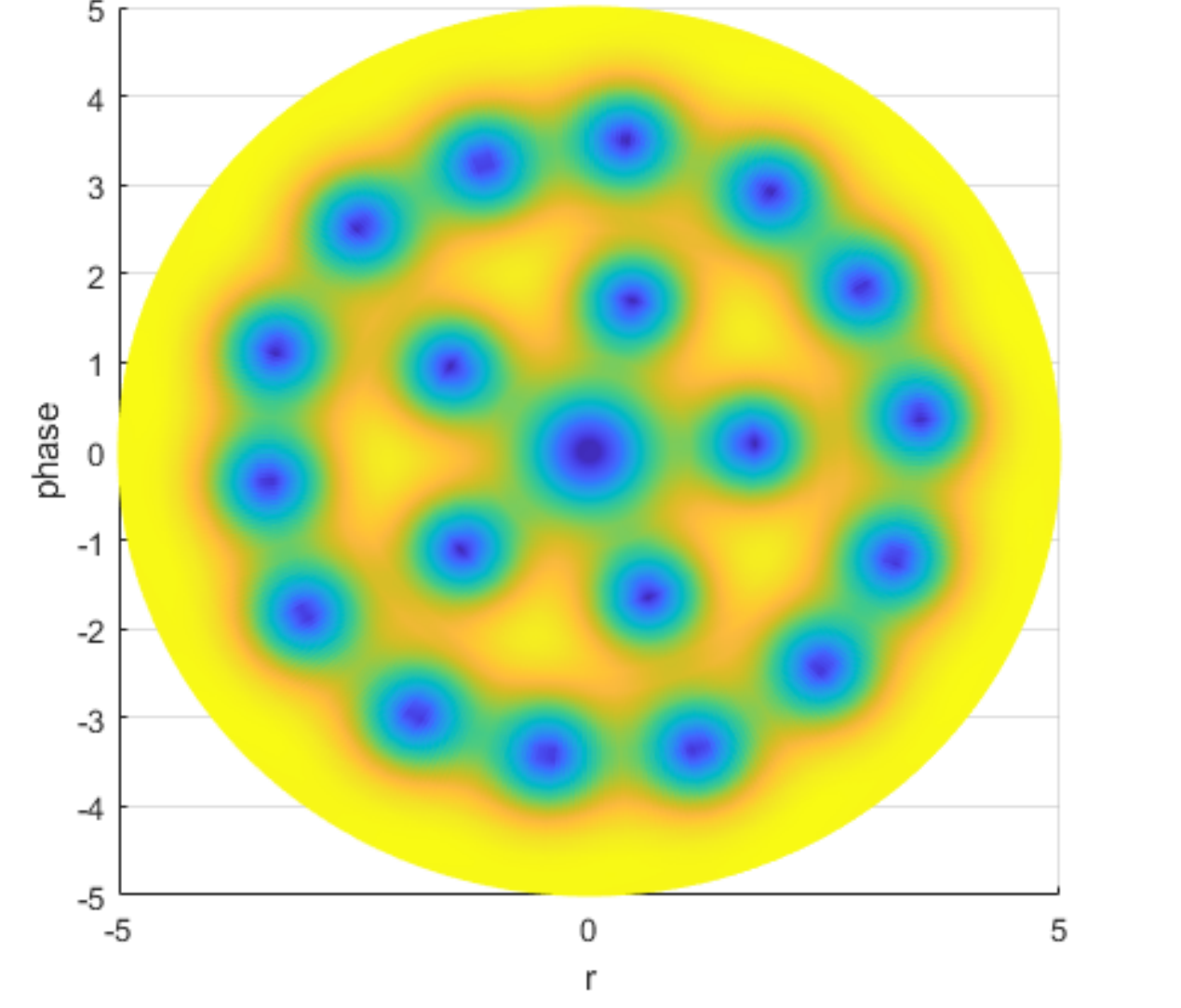}
\centerline{e}
\end{minipage}
\hspace{0mm}
\begin{minipage}{0.38\linewidth}
\includegraphics[trim=3.3cm 9.0cm 3.8cm 9.7cm, clip=true, scale=0.25, angle=0]{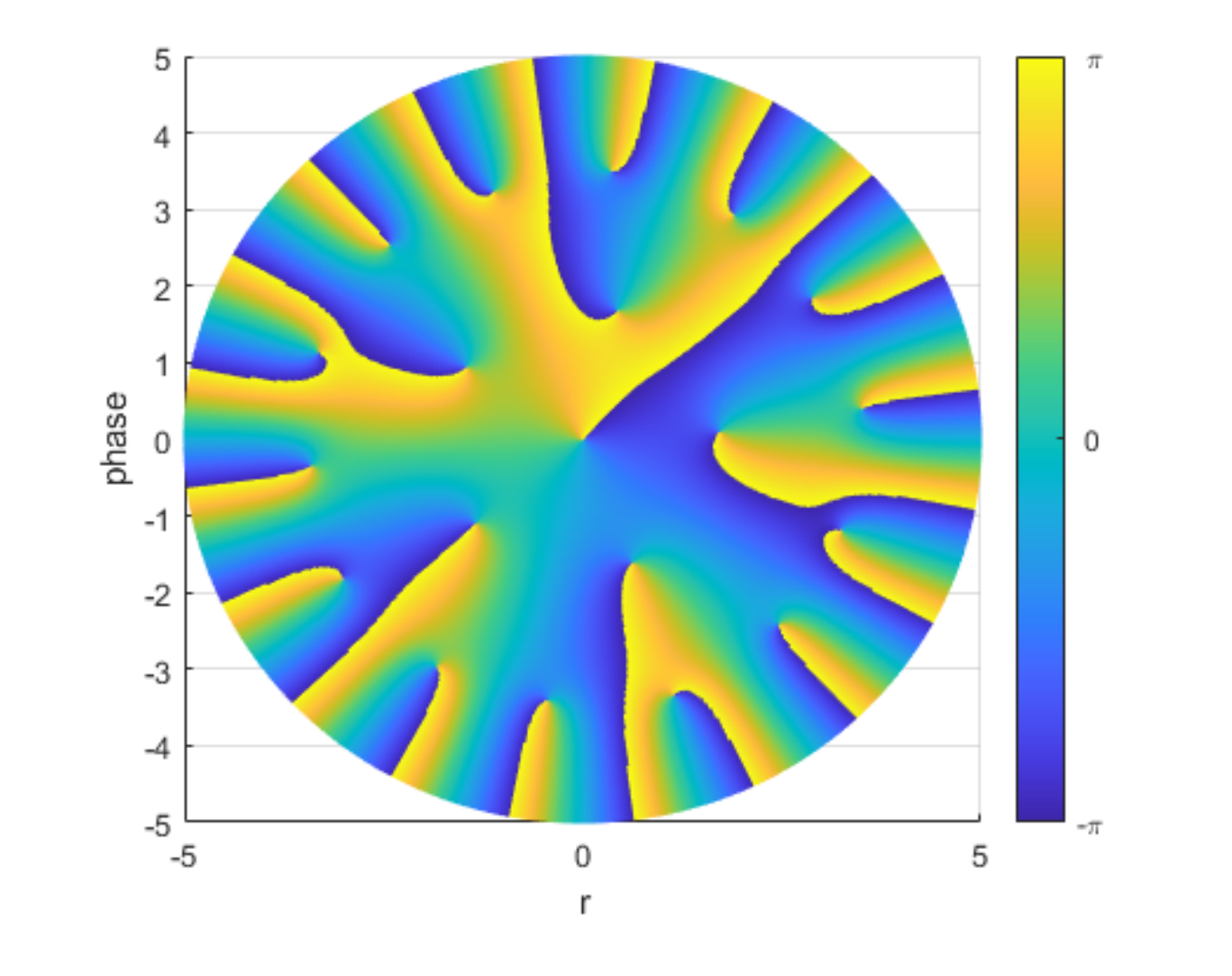}
\centerline{f}
\end{minipage}
\hspace{0mm}
\begin{minipage}{0.38\linewidth}
\includegraphics[trim=4.3cm 9.0cm 3.8cm 9.7cm, clip=true, scale=0.25, angle=0]{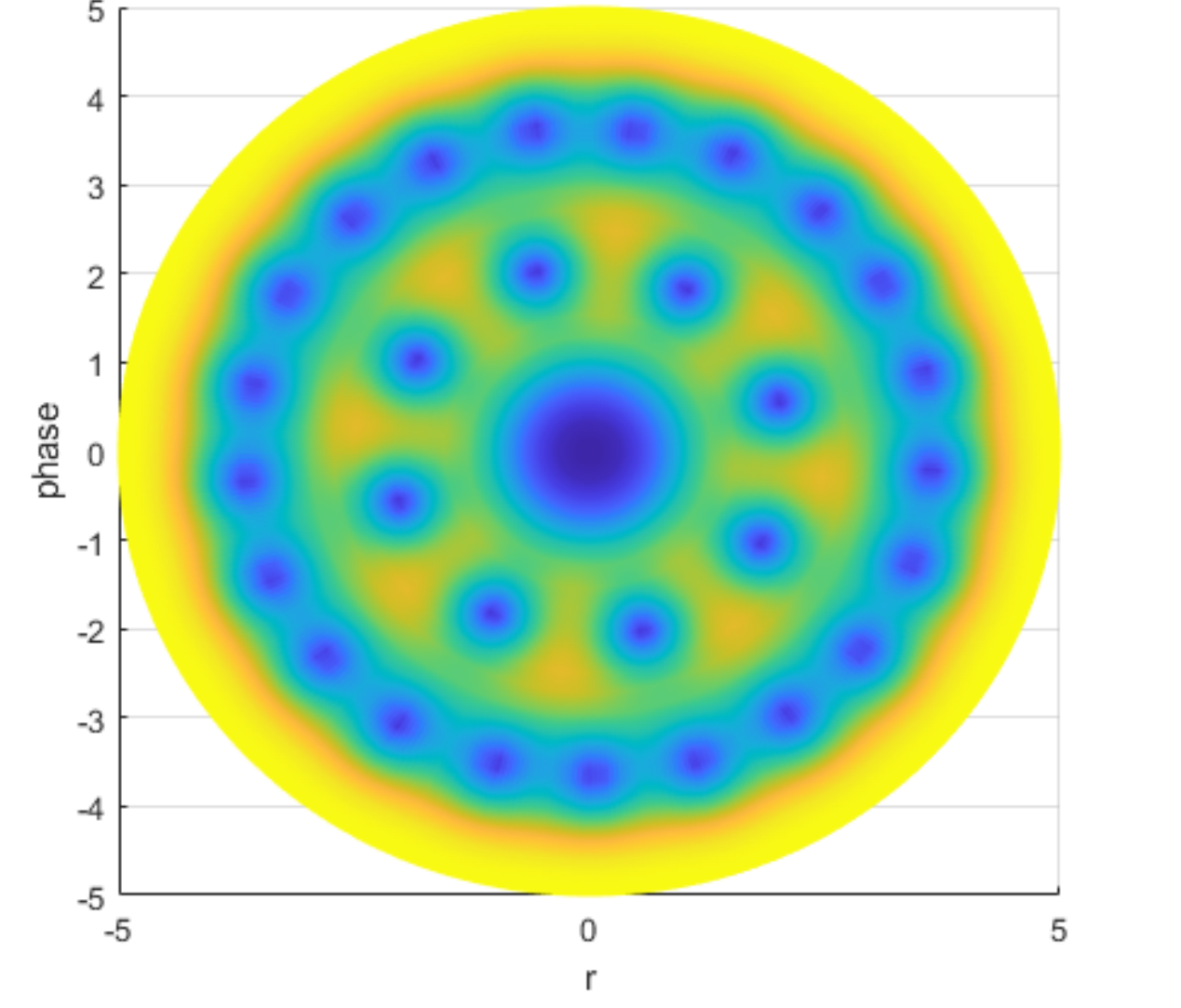}
\centerline{g}
\end{minipage}
\hspace{0mm}
\begin{minipage}{0.38\linewidth}
\includegraphics[trim=3.3cm 9.0cm 3.8cm 9.7cm, clip=true, scale=0.25, angle=0]{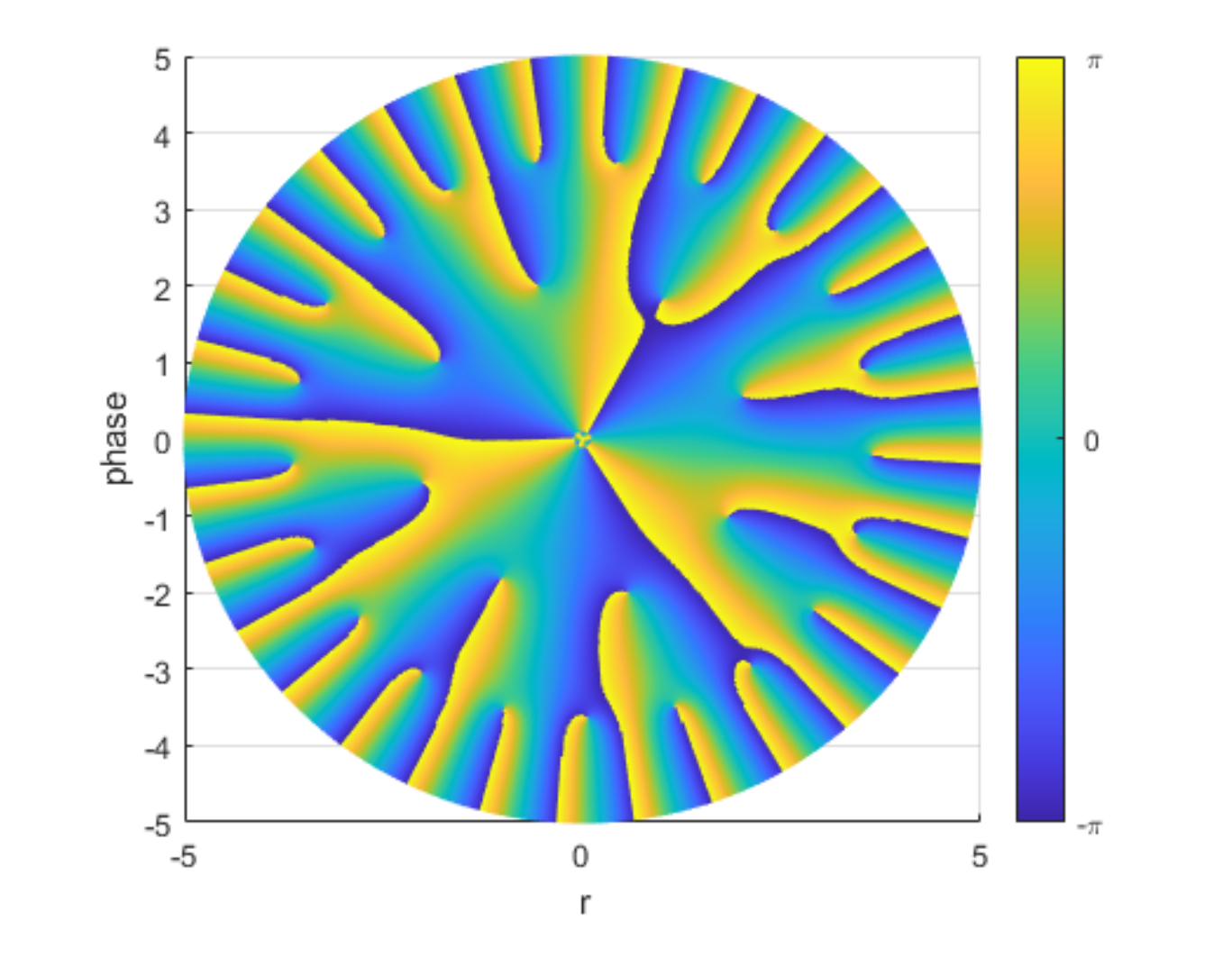}
\centerline{h}
\end{minipage}
\caption{\textbf{Order parameter (left) and phase (right) when $\Omega=0.5$ (a) (b), $\Omega=0.8$ (c) (d), $\Omega=1$ (e) (f), and $\Omega=1.5$ (g) (h)}.
  The value of the phase varies continuously from 0 (blue) to 2$\pi$ (yellow). When $\Omega=0.5$, a giant vortex whose winding number is two appears at the center of the dynamic system. But while $\Omega$ increased to 0.8, the giant vortex disappeared and split into four normal vortices form square structure, in which the phenomenon shows the giant vortex is not stable. In (e) (f) (g) (h), the giant vortex appears in two layer structure.}\label{fig2}
\end{figure}

In our simulation, after selecting fast rotations velocity beyond this velocity range, we observed more special vortex structures.
When the rotation velocity exceeds $0.45$, the vortices will enter the center of the hexagonal grid superfluid.
We cannot distinguish the number of charges in the disk center by  order parameters alone, so we must mention the diagram of phase configuration. In the diagram of phase configuration, the phase difference from blue to yellow is 2$\pi$, which corresponds to a single vortex.
As the velocity increases, there will be a large winding number vortex in the center of the holographic superfluid, whose winding number is 2n$\pi$ and n$>$1. It seems strange that, in a normal case, two vortices can combine together. But as shown in Fig.\ref{fig2}, we find that, in a fast-rotating disk, it is possible that multiple vortices gather at the center of the disk and form a giant vortex. The situation also can be obtained by using the G-P equation\cite{mine}.

Fig.\ref{fig2} (a) (b) shows that there are two charges in the center when $\Omega=0.5$, which form a giant vortex.
However, in Fig.\ref{fig2} (c) (d), where $\Omega=0.8$,
no giant vortex appear, only four inner layer vortices form a quartet lattice and eleven vortexes in the outer layer. This indicates that the formation of the giant vortex in the center requires not only sufficient rotation velocity, but also a specific value.

As the rotation velocity increases further, new physical images of holographic superfluid are constantly presented.
In Fig.\ref{fig2} (e) (f) (g) (h), we find a two-layer lattice structure in the disk. When the dynamic system has the rotation velocity $\Omega=1$, there is only one vortex core in the center of the disk.
An octagonal grid appears when $\Omega=1.5$ as shown in Fig.\ref{fig2} (g) (h). It can be seen that there are three vortices inside the octagonal grid, and they form a state of dynamic equilibrium due to mutual repulsion.
It is visible by naked eyes to observe at the location of center that three vortexes rise and fall. In this case, the theory of large wind instability has been fully verified. The theory mentions the vortices will mutually repel each other and tend to be independent.
We haven't seen the emergence of a three-layer structure yet, which might be found in  a larger disk.
Fig.\ref{fig2} (e) (f) and (g) (h) have similar structures, while in (g) (h), the faster case, the structure of the outer layer in the left picture has begun to blur, and there is a clear tendency to form a ring, but the vortex of the outer layer in this picture can still be distinguished. It can be counted that there are 21 vortices in the outer layer and the inner layer is octagonal grid. From the phase configuration on the right, it can be seen that the superfluid of the double lattice structure can also form a large winding vortex in the center when the rotation velocity is large enough.
From the above figure, we find that the phase configuration has a hierarchical structure, and each layer corresponds to a circle of vortices in the order parameter.

 Through Fig.\ref{fig2}, we discover the outermost phase difference is the total winding of the whole superfluid, the second layer phase difference includes the winding of the inner two layers, the phase difference of the center only shows the winding number of the giant vortex in the disk center. It can be summarized as a regular pattern:
The phase difference of the $\ N_{th}$ layer (it is stipulated that the number of layers from the inside to the outside is 1, 2, 3.... in turn) always contains the phase difference of all vortexes in the n-layer.

\begin{figure}[h]
\begin{minipage}{0.4\linewidth}
\includegraphics[trim=3.3cm 9.0cm 3.8cm 9.7cm, clip=true, scale=0.25, angle=0]{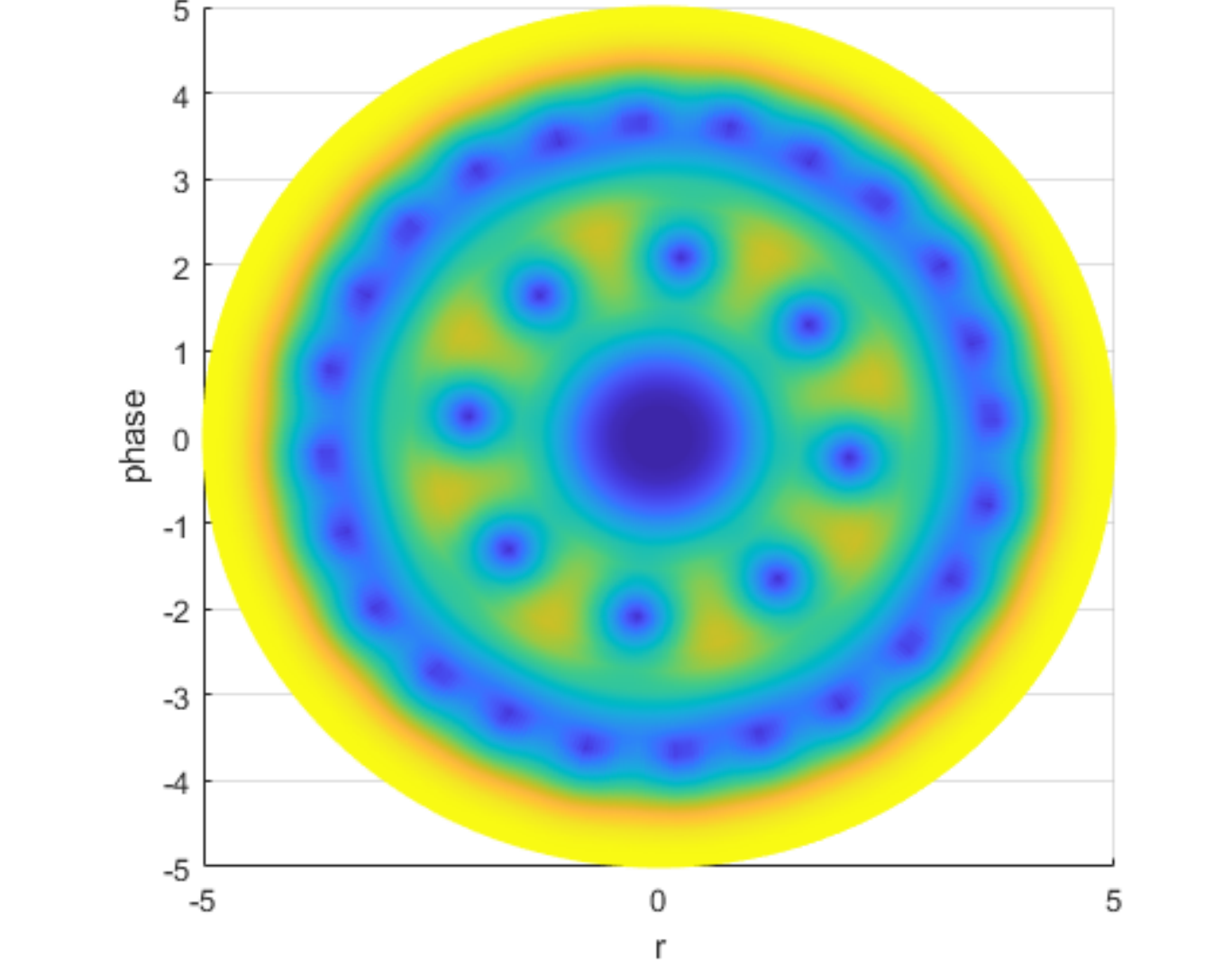}
\centerline{a}
\end{minipage}
\hspace{0mm}
\begin{minipage}{0.4\linewidth}
\includegraphics[trim=3.3cm 9.0cm 3.8cm 9.7cm, clip=true, scale=0.25, angle=0]{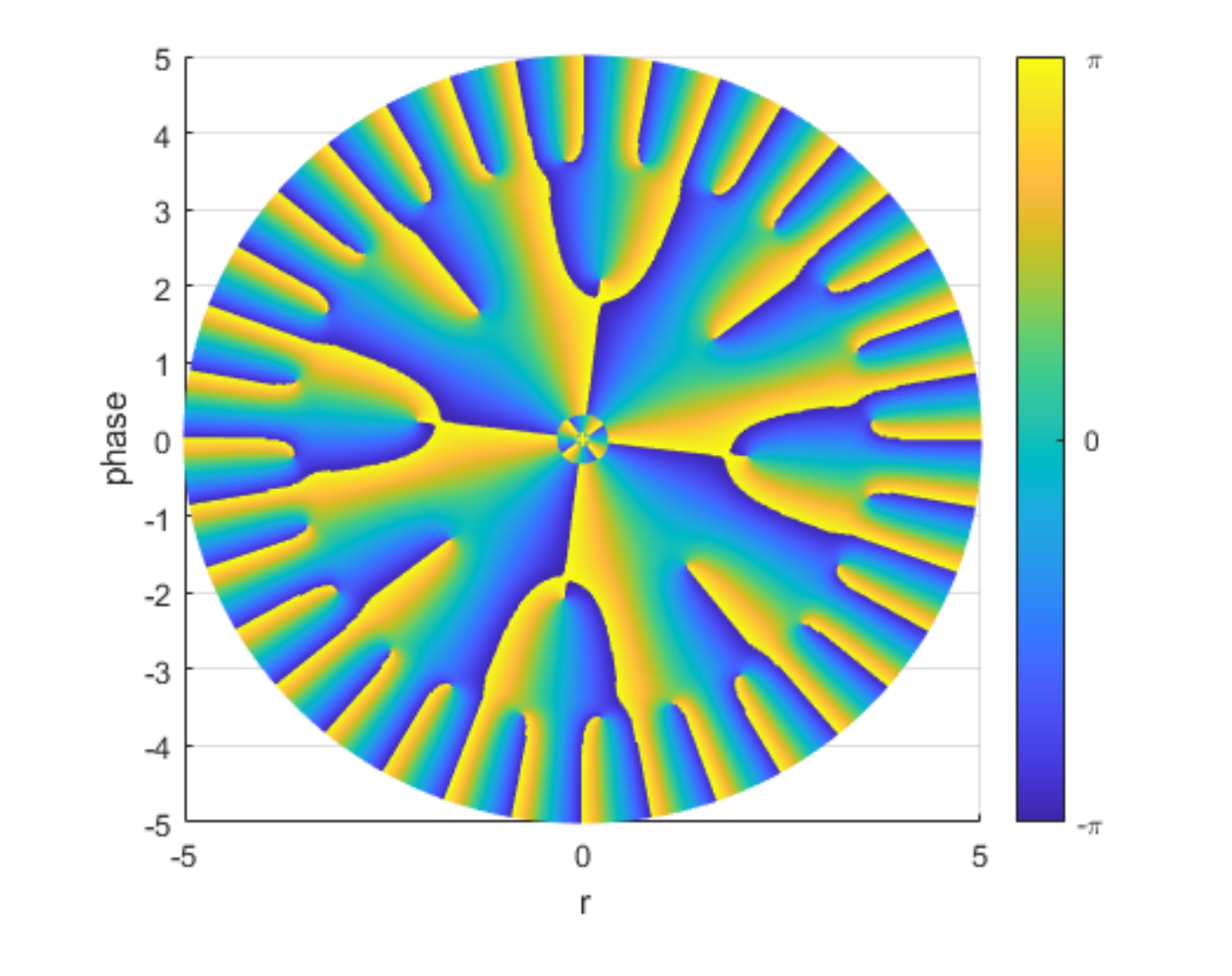}
\centerline{b}
\end{minipage}
\hspace{10mm}
\begin{minipage}{0.4\linewidth}
\includegraphics[trim=3.3cm 9.0cm 3.8cm 9.7cm, clip=true, scale=0.25, angle=0]{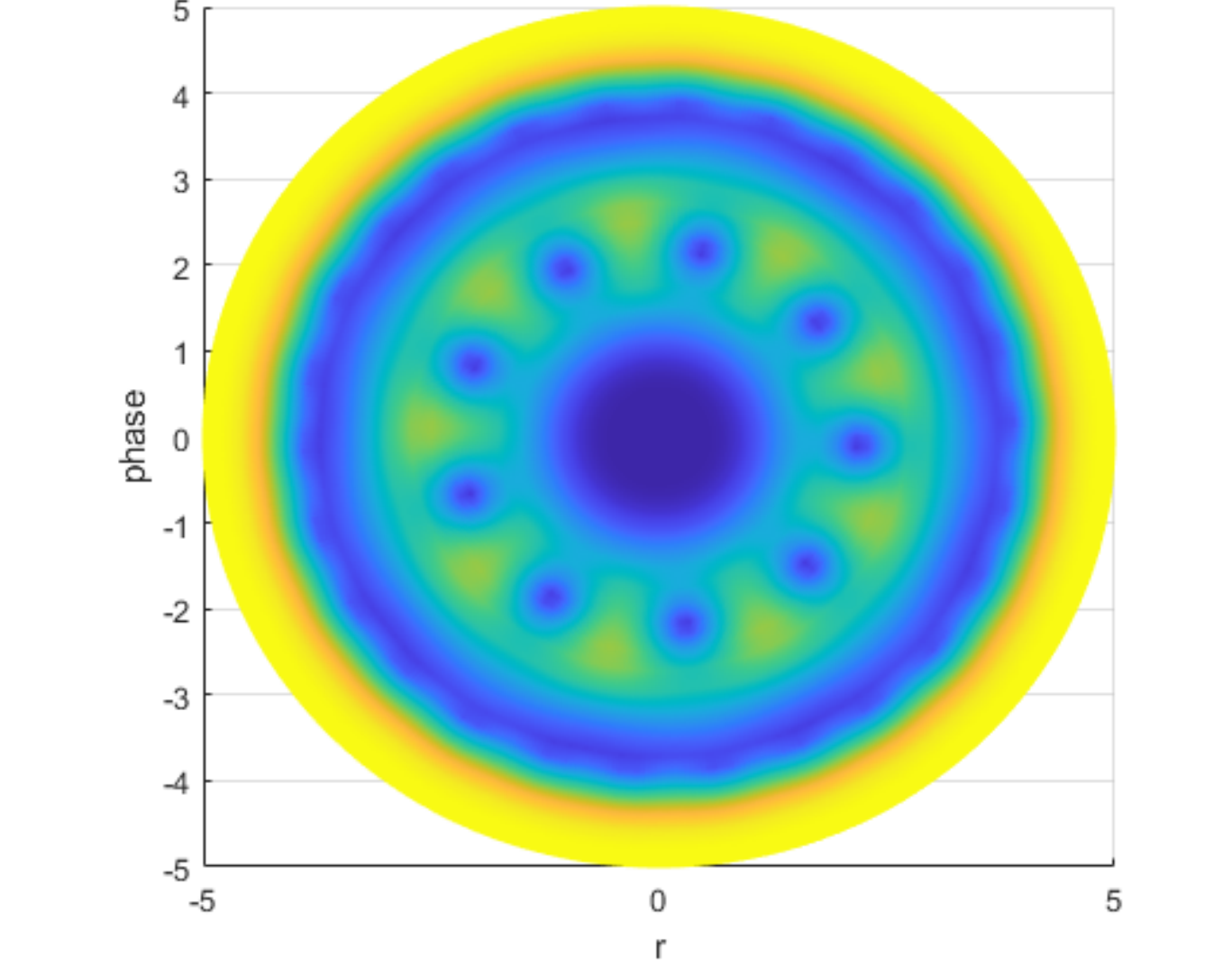}
\centerline{c}
\end{minipage}
\hspace{0mm}
\begin{minipage}{0.4\linewidth}
\includegraphics[trim=3.3cm 9.0cm 3.8cm 9.7cm, clip=true, scale=0.25, angle=0]{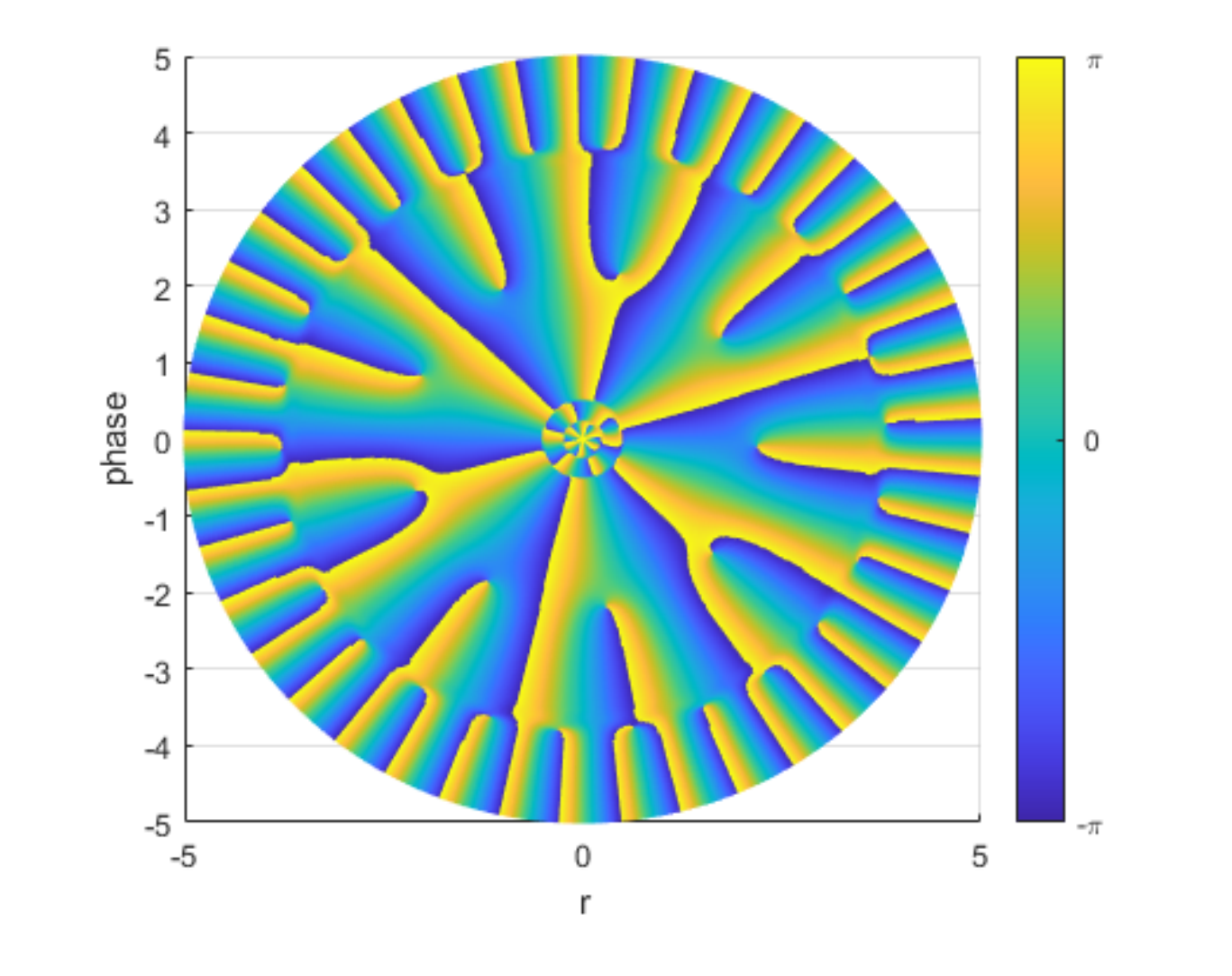}
\centerline{d}
\end{minipage}
\hspace{0mm}
\caption{ \textbf{Order parameter (left) and phase (right) when $\Omega=1.7$ (a) (b) and $\Omega=2$ (c) (d)}.
The phase configuration of giant vortex is stratified on both cases. We can count the winding number is 4 in (a) (b) and 6 in (c) (d).}\label{fig3}
\end{figure}

As the rotation velocity continues to increase, we have discovered a new giant vortex structure, which is different from the above giant vortices shown in Fig.\ref{fig1} and Fig.\ref{fig2} (a) (b) (g) (h).
The difference is not observable in the order parameter diagram, but it is clearly discernible from the diagram of phase configuration. We can see that in Fig.\ref{fig3}, when $\Omega=1.7$ and $\Omega=2$, there is just a difference at outer ring when we observe the order parameter, where the vortices at outer structure are completely a ring, but we can find that the structure of the giant vortex has changed in the diagram of phase configuration. When $\Omega=1.7$ for Fig.\ref{fig3} (a) (b), it is necessary to distinguish the number of winding in the diagram of phase configuration is 4 or 8. According to the above regular pattern from previous paragraph, it is easy to count how many charges are in a giant vortex. We mention that the phase differences of the second floor in the diagram of phase configuration (count from the outside to the inside) contains the giant vortex and next to giant vortex of phase differences, so the giant vortex contains four charges, it means the phase stratify at radius direction but no charge emerge in the disk center. And we can make sure the giant vortex on $\Omega=2$ for Fig.\ref{fig3} (c) (d) has 6 charges in the same way. Notice that the phase of the giant vortex in Fig.\ref{fig3} (b) is divided into two layers and in Fig.\ref{fig3} (d) divided into three layers, so stratification has no relation with the number of winding, just related to $\Omega$.

\begin{figure}[h]
\begin{minipage}{0.4\linewidth}
\includegraphics[trim=3.3cm 9.0cm 4.1cm 9.7cm, clip=true, scale=0.25, angle=0]{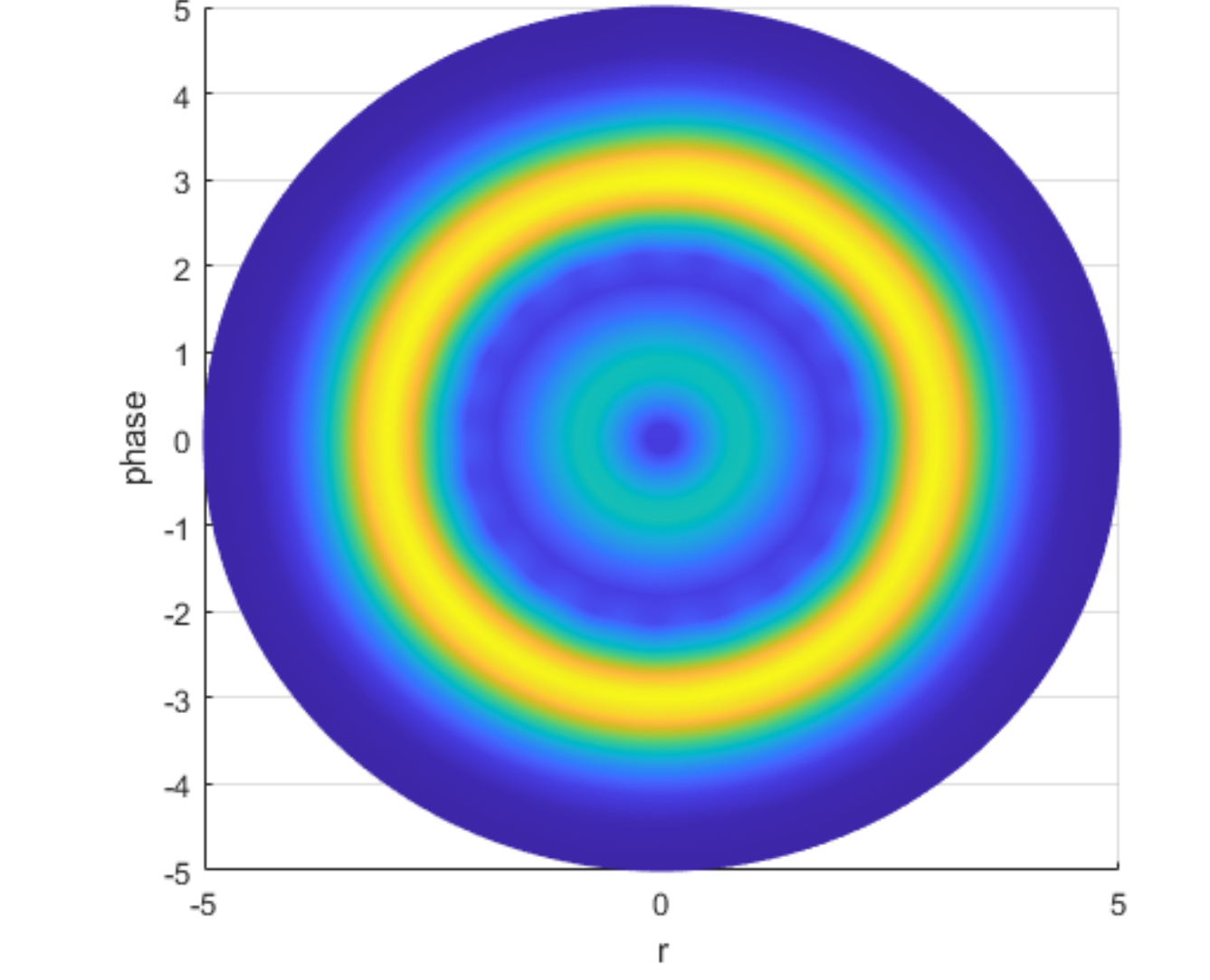}
\centerline{a}
 \end{minipage}
\hspace{0mm}
\begin{minipage}{0.4\linewidth}
\includegraphics[trim=3.3cm 9.0cm 3.8cm 9.7cm, clip=true, scale=0.25, angle=0]{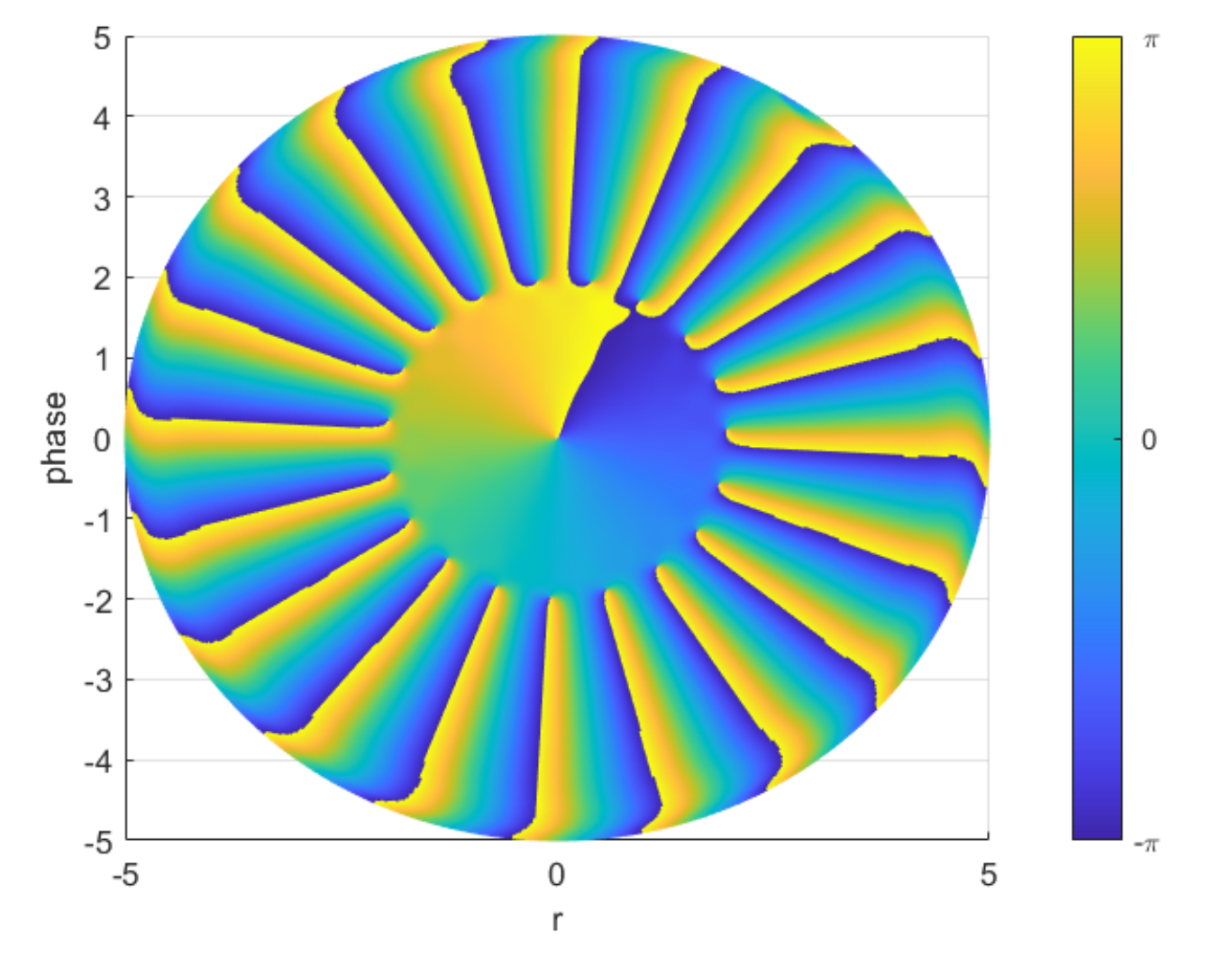}
\centerline{b}
 \end{minipage}
\hspace{0mm}
\caption{ \textbf{Order parameter (left) and phase (right) when $\Omega=2.5$ (a) (b)}.
The order parameter in (a) shows a superfluid ring and a vortex ring, the blue ring inside is the vortex ring, the yellow ring outside is the superfluid ring. The phase configuration is shown in (b).}\label{fig4}
\end{figure}

In Fig.\ref{fig4} with $\Omega=2.5$, we find a superfluid ring outside and a vortex ring inside, which is similar to the result of the G-P equation\cite{gpt}. When $\Omega$ is extremely fast, the vortex structure will be destroyed, the phase will be chaos, so that is not in consideration.

\section{THE RELATIONSHIP BETWEEN WINDING NUMBER AND ROTATING VELOCITY}

\begin{figure}[ht]
\centering
\includegraphics[trim=3.3cm 9.0cm 1.0cm 8.7cm, clip=true, scale=0.6, angle=0]{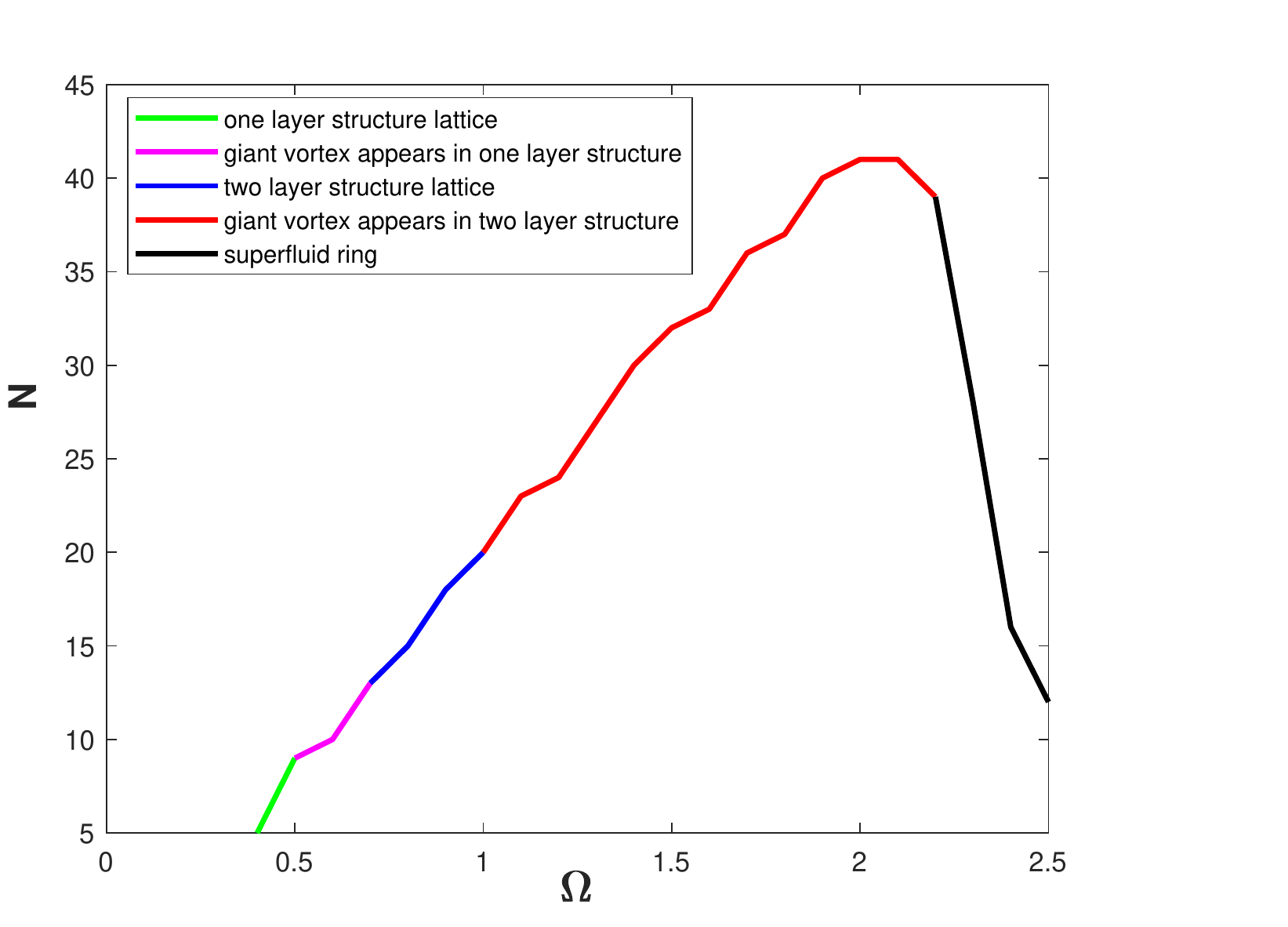}
\caption{\textbf{Relationship between quantity of vortex N and rotating velocity $\Omega$}. Green line on behalf of the one layer structure lattice state. The concrete physics picture of the green line is given in Ref\cite{Z}. The magenta line represents the state of giant vortex appears in one-layer structure. The blue line represents the two-layer structure lattice state which no giant vortex arises. The red line represents the state of giant vortex appears in two-layer structure. The black line represents the superfluid ring state.}\label{fig5}
\end{figure}

As $A_\theta$ increases, more vortices can enter the disk. here is a conjecture: if the $\Omega$ is large enough, a lot of vortices will emerge in the fixed size disk, a vortex is easy to interact with others vortices close to it, so the way that merge a giant vortex in the disk centre to make the system stable. This kind of state is corresponds the magenta and red line region in Fig.\ref{fig5}.
The situation of Fig.\ref{fig2} (a) (b) is contained in the magenta line region, and Fig.\ref{fig1}, Fig.\ref{fig2} (g) (h) and Fig.\ref{fig3} (a) (b) (c) (d) are contained in the red line region. We mentioned that there is no giant vortex state in the blue line region which corresponds with Fig.\ref{fig2} (c) (d) (e) (f), because the system forms two layer structure from one layer structure lattice in the rotational velocity. So vortices won't all squeeze into the middle of the disk centre when the vortex quantity is not very big in the dynamics progress. According to previous theories and results, the charge should rise if the $\Omega$ is extremely fast, the fact is, the quantity of vortex is decreasing when the rotating velocity $\Omega$ keeps increasing, as shown in the black line region in Fig.\ref{fig5}, because the order parameter is extremely suppressed. Superfluid ring appears in the region. In the front four region, the quantity of vortex and rotating velocity approximately yield the Feynman relation\cite{Feynman}, but as the increasing quantity of vortex in the red region, the effect of dissipation between vortex and superfluid influences the Feynman relation obviously.

\section{SUMMARY}
We have numerically simulated the quantum multiple body system by applying a holographic model on the fast rotating superfluid under the finite temperature. We have found that the unusual formation giant vortex exists stable in the holographic superfluid due to the fast rotating disk, the giant vortex contains multiple charges and emerges the phase configuration stratification on the radius direction. We also found the formation of a superfluid ring and a vortex ring exist stable in the holographic superfluid due to the ultra-fast rotating disk, this formation and the phase configuration are undiscovered before. As the rotating velocity increases, there are five region in the over-all dynamics process shown in Fig.\ref{fig5}. The first region of the process is a one-layer vortex lattice. The second region is that a giant vortex appears at the system centre based on a one-layer lattice. The third region is two-layer vortex lattice. The four region is that a giant vortex appears at the system centre based on a two-layer lattice. In particular, the phase of the giant vortex will stratify in region four. The fifth region is a superfluid ring and a vortex ring. The quantity of vortex increases approximately linearly with angular velocity, .i.e Feynman relation, until $\Omega\approx$2. When $\Omega>2.1$, the number  of vortices will decrease rapidly with increasing angular velocity, the order parameter is extremely suppressed and forms a superfluid ring and a vortex ring on the ultra-fast rotating disk.

\begin{acknowledgments}
This work is supported by the National Natural Science
Foundation of China (under Grants No. 11275233).
\end{acknowledgments}

\vspace{0mm}

%\bibliography{ref}


\end{document}